\documentclass[letterpaper,12pt,twoside]{uncthesis}
\uncdegreename{Doctor of Philosophy}
\uncthesisdepartment{Department of Computer Science}
\uncthesistype{dissertation}

\uncthesistitle{Mining Secure Behavior of Hardware Designs}
\uncthesisauthor{Calvin Deutschbein}
\uncthesisuniversity{University of North Carolina at Chapel Hill}
\uncthesisyear{2021}
\uncthesisadvisor{Cynthia Sturton}
\uncthesiscommittee{Cynthia Sturton\\
Peng Huang\\
Fabian Monrose\\
Donald Porter\\
Michael Reiter
}

\usepackage{times}
\begin{document}
\pagenumbering{roman}
\newgeometry{left=\uncthesisleftmargin in,top=2in,right=\uncthesisrightmargin in,bottom=1in,nohead}

\begin{titlepage}

\begin{singlespace}
\centering

\currentpdfbookmark{TITLE}{bk:title}

\vspace{1in}
\begin{adjustwidth}{0.5in}{0.5in}
\centering
\MakeUppercase{\uncthesistitle}
\end{adjustwidth}

\nointerlineskip\vspace{1in}

\uncthesisauthor{}

\nointerlineskip\vspace{1in}

\noindent
A \uncthesistype{} submitted to the faculty at the \uncthesisuniversity{} in partial fulfillment of the requirements for the degree of \uncdegreename{} in the \uncthesisdepartment{}.

\nointerlineskip\vspace{1in}

Chapel Hill\\
\uncthesisyear{}

\end{singlespace}


\nointerlineskip\vspace{0.71in}
\begin{flushright}
\begin{minipage}{2.1in}
\setlength{\tabcolsep}{0em}
\begin{tabular}{l}
Approved by: \\
\uncthesiscommittee{}
\end{tabular}
\end{minipage}
\end{flushright}

\end{titlepage}

\newgeometry{left=\uncthesisleftmargin in,top=8.42in,right=\uncthesisrightmargin in,bottom=1in,nohead}

\begin{center}
\begin{singlespace}
\copyright~\uncthesisyear\\
\uncthesisauthor \\
ALL RIGHTS RESERVED
\end{singlespace}
\end{center}

\clearpage
\newgeometry{left=\uncthesisleftmargin in,top=2in,right=\uncthesisrightmargin in,bottom=1in,nohead}
\restoregeometry

\begin{abstract}
\addtolength{\topmargin}{-1.0in}
Hardware presents an enticing target for attackers attempting to gain access to a secured computer system. Software-only exploits of hardware vulnerabilities may bypass software level security features. Hardware must be made secure. However, to understand whether a hardware design is secure, security specifications must be generated to define security on that design. Micro-architectural design elements, undocumented or under-documented features, debug interfaces, and information--flow side channels all may introduce new vulnerabilities. The secure behavior of each must be specified in order ensure the design meets its security requirements and contains no vulnerabilities. However, manual efforts can be overwhelmed by design complexity, and many hardware vulnerabilities, such as Memory Sinkhole, SYSRET privilege escalation, and most recently Spectre/Meltdown, persisted in product lines for decades despite extensive testing. An automated solution is needed to specify secure designs.
\addtolength{\topmargin}{1.0in}
Specification mining offers a solution by automating security specification for hardware. Specification miners use a form of machine learning to specify behaviors of a system by studying a system in execution. However, specification mining was first developed for use with software. Complex hardware designs offer unique challenges for this technique. Further, specification miners traditionally capture functional specifications without a notion of security, and may not use the specification logics necessary to describe some security requirements.

This work demonstrates specification mining for hardware security. On CISC architectures such as x86, I demonstrate that a miner partitioning the design state space along control signals discovers a specification that includes manually defined properties and, if followed, would secure CPU designs against Memory Sinkhole and SYSRET privilege escalation. For temporal properties, I demonstrate that a miner using security specific linear temporal logic (LTL) templates for specification detection may find properties that, if followed, would secure designs against historical documented security vulnerabilities and against potential future attacks targeting system initialization. For information--flow hyperproperties, I demonstrate that a miner may use Information Flow Tracking (IFT) to develop output properties containing designer specified information--flow security properties as well as properties that demonstrate a design does not contain certain Common Weakness Enumerations (CWEs).

\end{abstract}
\renewcommand{\contentsname}{\hfill\bfseries\normalsize TABLE OF CONTENTS\hfill}
\renewcommand{\cftaftertoctitle}{\hfill}
\renewcommand{\cftdotsep}{1.5}
\cftsetrmarg{1.0in}

\setlength{\cftbeforetoctitleskip}{56pt}
\setlength{\cftaftertoctitleskip}{18pt}

\renewcommand{\cftchapfont}{\normalfont}
\renewcommand{\cftchappagefont}{\normalfont}
\renewcommand{\cftchapleader}{\cftdotfill{\cftdotsep}}

\setlength{\cftbeforechapskip}{15pt}
\setlength{\cftbeforesecskip}{10pt}
\setlength{\cftbeforesubsecskip}{10pt}
\setlength{\cftbeforesubsubsecskip}{10pt}

\begin{singlespace}

\currentpdfbookmark{TABLE OF CONTENTS}{bk:contents}
\tableofcontents
\end{singlespace}

\clearpage

\renewcommand{\listtablename}{LIST OF TABLES}
\phantomsection
\addcontentsline{toc}{chapter}{LIST OF TABLES}

\setlength{\cftbeforelottitleskip}{-14pt}
\setlength{\cftafterlottitleskip}{22pt}
\renewcommand{\cftlottitlefont}{\hfill\normalsize\bfseries}
\renewcommand{\cftafterlottitle}{\hfill}

\setlength{\cftbeforetabskip}{12pt}
\cftsetrmarg{1.0in}

\setlength{\cfttabindent}{0in}

\begin{singlespace}
\listoftables
\end{singlespace}

\clearpage

\renewcommand{\listfigurename}{LIST OF FIGURES}
\phantomsection
\addcontentsline{toc}{chapter}{LIST OF FIGURES}

\setlength{\cftbeforeloftitleskip}{-14pt}
\setlength{\cftafterloftitleskip}{22pt}
\renewcommand{\cftloftitlefont}{\hfill\normalsize\bfseries}
\renewcommand{\cftafterloftitle}{\hfill}

\setlength{\cftbeforefigskip}{12pt}
\cftsetrmarg{1.0in}

\setlength{\cftfigindent}{0in}

\begin{singlespace}
\listoffigures
\end{singlespace}

\clearpage

\phantomsection
\addcontentsline{toc}{chapter}{LIST OF ABBREVIATIONS}

\begin{center}
\textbf{LIST OF ABBREVIATIONS}
\vspace{16pt}
\end{center}


\noindent
\begin{tabular}{p{0.8in} p{5in}}
ACM     & Access Control Module \\
CISC    & Complex Instruction Set Computer \\
CPU     & Central Processing Unit\\
CWE     & Common Weakness Enumeration\\
HDL     & Hardware Design Language \\
IFT     & Information Flow Tracking \\
ISA     & Instruction Set Architecture \\
LTL     & Linear Temporal Logic\\
OS      & Operating System\\
RISC    & Reduced Instruction Set Computer\\
RTL     & Register Transfer Level \\
SMM     & System Management Mode \\
SoC     & System on a Chip\\
TLM     & Transcation Level Model \\
VCD     & Value Change Dump\\
\end{tabular}

\clearpage

\pagenumbering{arabic}

\chapter{Introduction}
\label{chap:intro}

Hardware presents an enticing target for attackers attempting to gain access
to a secured computer system. 
Software-only exploits of hardware vulnerabilities may bypass software level 
security features. Hardware must be made secure. 
However, for designers to determine whether hardware is secure, security 
specifications must be generated for a given design.
Micro-architectural design elements, undocumented or under--documented features, debug 
interfaces, and information--flow side channels all may introduce new 
vulnerabilities. 
The secure behavior of each must be specified in order to ensure the design 
meets its security requirements and contains no vulnerabilities. 
Yet manual efforts can be overwhelmed by design complexity.
An automated solution is needed to specify secure designs.
Mining secure behavior for hardware designs offers an automated approach to security
specification, an important step 
toward achieving secure hardware design.

In hardware designs, weaknesses or vulnerabilities may be
introduced, usually alongside new design features offering higher performance,
that persist within product lines for years or decades. 
Consider the Memory Sinkhole vulnerability~\citep{Domas2015Sinkhole}. Introduced to x86 
designs in the mid-1990’s with the new System Management Mode (SMM), it remained
exploitable on x86 architectures until the Sandy Bridge-EP release in 2011.
Despite affecting hardware for almost two decades, the vulnerability 
was first publicly demonstrated only in 2015.
The vulnerable hardware was intended to be secure, and tested for security, 
but was not validated against a security specification that precisely disallowed 
the Memory Sinkhole's attack mechanisms. In Chapter~\ref{chap:Astarte}, I show
how mining can automatically generate such a security specification.

Mining can also promote best practice for the development of secure designs. 
For example, the
Common Weaknesses Enumerations (CWEs) database describes high level design goals
for secure systems, often for software. 
Recently, as part of an industry-wide effort to secure 
hardware, CWEs describing many forms of secure design for hardware have been
added to the database as well.
CWEs may be specified over particular types of designs and represent best practices
for generally secure design, such as debug and reset implementations. 
While assessing designs against CWEs relies on manual efforts, including
reasoning about what security means for a given design, CWEs may offer an early 
line of defense against broad spectrums of attacks.
Automated application of CWEs to designs could create designs secure against
vulnerabilities to potential future attacks by
eliminating common weaknesses. In Chapter~\ref{chap:Isadora}, I show
how mining can automatically generate a specification containing CWE--relevant
properties for a given design.


Specification mining offers a powerful tool that designers may use
as part of broader efforts to prevent future attacks.
For the many designs for which there is no 
existing security specification, security
specification mining can generate one.
These generated specifications may then be used for
validation efforts.
Specification miners use a form of machine learning to specify behaviors of a
system by studying the execution of said system, and these behaviors may include the 
secure operations of intentionally secure designs.
In this work, I show how specification mining can address challenges of 
developing security specifications for different designs, ranging from
x86 to SoC designs, and for different
logics of specifications, including linear temporal logic (LTL) properties
and hyperproperties.
I design and implement a series of specification miners to create security 
specifications
of hardware, and evaluate generated
specifications against manually defined notions of security. Each
miner uses unique methods developed for hardware security mining to
define specific security goals.

Specification miners consider a design by taking as input traces of execution
and producing as output lists of 
properties of the design.
While traces
may take different forms, they capture in some way the state of a 
design at different time points during a run of execution, such as in
 value change dumps (VCDs), which log every time
the value held by a register or wire changes, or as in debug logs, which may log
software-visible signals across clock cycles or control flow changes. 
Specification miners may produce invariants, logical expressions that must hold
over some design elements at all time points within a set of traces.
Miners may also consider temporal properties, such as properties specified in 
Linear Temporal Logic (LTL), which define behavior across different time points. 
Miners may even consider hyperproperties, which require reasoning across 
multiple runs of execution of a design and therefore require reasoning about 
more than a single trace.
Increasingly complex logics of specification allow defining security properties
preempting correspondingly complex attack patterns.
A temporal property may be used to secure a design against timing attacks, or a 
hyperproperty may be used to secure a design against information flow leakage.

Specification mining was pioneered for use on software, and
state--of--the--art specification miners are often 
not intended for use on hardware designs.
Early specification miners, such as DIDUCE~\citep{Hangal2002Diduce}, which
studied Java programs,
or Perracota~\citep{Yang2006Perracotta}, which studied APIs,
 were specifically based
around common errors in programs.
Miners developed for hardware designs, which lack abstractions present in 
software such as types or 
functions, used different approaches to understand design behavior.
Earlier hardware miners, such as IODINE~\citep{Hangal2005Iodine}
and later on, GoldMine~\citep{Hertz2013GoldMine}, 
extracted only invariants, and often only specific kinds of invariants,
such as one--hot encoding. These invariants were valuable for 
studying designs, but were not intended to capture all security agreements
of the underlying hardware.
Later hardware miners using more descriptive logics, such as 
SAM~\citep{Li2010Scalable} and
A-TEAM~\citep{Danese2017ATEAM} with temporal logics
and Hyperminer~\citep{Rawat2020Hyperminer} with hyperproperties,
generate valuable specifications more expressive than earlier tools.
However, challenges remained to automatically generate properties 
such as temporal expressions
over equalities between registers or information flows
that may occur but only under certain conditions.
And as expressiveness improves, mining hardware designs may produce far more 
properties than can be reasonably considered by human designers.
 
Mining for security introduces further challenges, as specification
miners do not innately differentiate security and functional correctness.
Especially when considering designs for which no security specification exists,
identifying which properties enact some 
implicit security agreement may require domain 
specific or even design specific solutions. 
Even when specifications cover some cases of 
insecure behavior identified on a given design, it is another
matter entirely to compose these
disparate examples into a notion of design--level security, much less 
a method for understanding secure design generally.
Specification mining for security must engage with the question of 
what is, and is not, a security property.

Secure hardware behavior offers an ever-moving target for 
security researchers and hardware designers.
This work must contend with what it means for a property to be a security
property, or a specification to be a security specification.
In the case of known attacks, such as Memory Sinkhole~\citep{Domas2015Sinkhole}, 
security properties may be understood to be some specification mining
output property that, if followed, would mean the attack could not be used
against the design.
Yet secure behavior with respect to known attacks does not necessarily
mean a design is secure against all possible attacks.
Vulnerabilities may only be classified as vulnerabilities after some attack
is discovered, and many designs do not have complete functional specifications that are
validated for correctness or complete enumerations of their security requirements
in order to anticipate what forms an attack may take.
Consequently, generated security specifications
can, and in many cases should, contain properties for which there is no known attack or exploit.
Further, secure behavior for hardware
designs must also consider expectations of software, 
including operating systems,
that run on the design. 
Hardware features, such as memory protections or privilege levels,
may be implemented to service software-level requirements.
When automatically generating specifications with the 
intent to support the security validation process, an automated process may target:
\begin{itemize}
\item Software expectations of hardware for security
\item Historical examples of exploits
\item Existing best practices
\end{itemize}
Each of these cases are considered in this thesis, often in conjunction.

While scaling to hardware and refining to security both represent challenges,
specification mining remains well placed to specify hardware security.
In fact, the additional complexities of hardware can in part be addressed
by considering security cases specifically, and the comparative complexity
itself makes use of automated tools more desirable. 
The usefulness of 
specification mining increases further when considering complex notions
of security such as temporal properties or hyperproperties. 
This thesis demonstrates that specification mining can solve vital challenges for
secure hardware designs by exploring three cases where security specifications
would assist designers in securing hardware.

To demonstrate security specification mining on Complex Instruction Set Computer
(CISC) architectures, I created
Astarte, a specification miner that partitions the x86 design 
state space along control signals that govern secure behavior of the processor.
Astarte works at the Instruction Set Architecture (ISA) level and may therefore 
be used for hardware designs that are closed source.
By evaluating the output specification against security properties from a 
manual review of design documentation and additional properties
capturing correct design behavior with respect to two historical attacks, 
Memory Sinkhole~\citep{Domas2015Sinkhole} and SYSRET privilege 
escalation~\citep{Xen2012SYSRET}, I show that the Astarte specifications contain properties 
that are security relevant in each of these cases.
Astarte addresses software expectations of hardware for security by considering
how operating systems interface with underlying hardware. Some hardware features,
such as privilege levels and operating modes, are assumed to provide certain controls,
and software may make assumptions of how underlying hardware manages internal
state that introduce potential exploits. 
This was the case for SYSRET privilege escalation, where the SYSRET
instruction had different specified behaviors on AMD and Intel designs. When 
software assumed the AMD behavior applied to Intel
designs, the unanticipated hardware behavior allowed user--level attackers to 
elevate their privilege level.
By applying Astarte to traces generated by different operating systems
and comparing output properties across them,
Astarte may provide insight into the security expectations software
may have of hardware.
In turn, hardware designers may implement or preempt design behavior
related to
these expectations to achieve the assumed notions of security.
Astarte is presented in Chapter~\ref{chap:Astarte} and is
based on joint work with Cynthia Sturton~\citep{deutschbein2020astarte}.

To demonstrate security specification mining of temporal properties, I created
Undine, a security specification miner that uses security 
specific LTL templates to mine security properties.
Undine targets Reduced Instruction Set Computer
(RISC) central processing unit (CPU) or System on a Chip (SoC) designs.
As these designs are open source, Undine works at the Register Transfer Level (RTL).
Undine is situated as an extension of the invariant detection methods of SCIFinder~\citep{Zhang2017SCIFinder} to temporal properties, and similarly is oriented
toward describing secure design behavior against 
historical examples of exploits while also producing new properties.
The output specification provides properties that, if followed, would prevent 
known and potential 
future attacks on hardware that can be defined before, after, and across 
system state transitions.
I show an example exploit of privilege escalation that violates a generated
temporal
property specifying system behavior prior to reset in Section~\ref{sec:exploit}.
Undine uses a labelling system of events to generate LTL properties
over propositional variables including equalities between registers and
over registers and values for RISC CPUs.
Doing so results in output specifications with high coverage and density 
of historically established security properties.
To produce these properties, Undine uses 
a library of labelled linear temporal logic
templates useful across multiple designs, a valuable product of this
line of research.
Undine is presented in 
Chapter~\ref{chap:Undine} and is based on joint work with Cynthia 
Sturton~\citep{deutschbein2018undine}.



To demonstrate security specification mining of information flow
properties, I created Isadora, a security specification miner that uses
information flow tracking to mine information flow specifications.
Like Undine, Isadora works at the Register Transfer Level (RTL) and is suitable 
for use on open source hardware designs including RISC CPUs and SoC designs.
The output specification gives the information flow relations between all design 
elements, specifying whether flows may occur between two elements and, if so, 
specifying the design conditions under which information flow occurs.
By evaluating the output specification against designer--provided sets of security
properties and against high level Common Weakness Enumerations (CWEs), 
I show that the Isadora miner finds security properties representing 
security efforts of individual designers for a specific design 
and security properties representing established best practices for
security.
The evaluation considers multiple
designs, including an access control module, an SoC design, and a RISC-V CPU.
This is presented in Chapter~\ref{chap:Isadora} and is based on joint work 
with Andres Meza, Francesco Restuccia, Ryan Kastner, and Cynthia Sturton
~\citep{Deutschbein2021Isadora}.

The thesis of this work is:
\begin{quote}
Specification mining can discover 
properties that can be used to verify the secure behavior of 
closed source CISC CPU designs, 
properties that can be used to verify the 
temporal correctness of CPU designs, and 
hyperproperties that can be used to verify that 
modules, SoCs, and CPUs have secure information flow.
\end{quote}

The techniques developed and presented in this thesis can enable hardware
designers to better specify their designs with respect to security, either
fully automatically or alongside existing security efforts, and reduce the 
barriers to developing more secure hardware.
\chapter{Related Work}
\label{chap:rw_full}

I organize work related to this thesis 
using three dichotomies:
\begin{itemize}
\item Manual versus Automatic
\item Software versus Hardware 
\item Functional Correctness versus Security
\end{itemize}

\section{Automatic Software Functional Correctness} 

Specification mining as a technique was introduced in~\cite{Ammons2002mining} 
in which execution traces are examined to infer
protocol specifications in the form of regular expressions. 
~\cite{Weimer2005mining} used
both static and dynamic traces to filter out less useful candidate
specifications.
The Perracotta miner~\citep{Yang2006Perracotta} tackled the
challenges posed by having imperfect execution traces
and by the complexity of the search space.
DIDUCE~\citep{Hangal2002Diduce} studied software designs by instrumenting
them to extract invariants online rather than offline.
The Daikon Dynamic Invariant Detector~\citep{Ernst2007Daikon} learns
properties that express desired semantics of a program running offline
on execution traces. 

While described as intended to discover program invariants, 
Daikon represents an ongoing and multi-decade research effort
in dynamic invariant detection, and its underlying inference 
engine offers a powerful tool for exploration
of hardware designs as well.
For this reason, 
in this work Daikon is used within
both the Astarte and Isadora framework. Astarte applies Daikon in multiple
passes over trace sets to produce properties conditioned on extracted
control signals. Isadora applies Daikon to trace slices identified by
information flow tracking to define predicates that specify
design conditions during information flows.

Specification mining has also been applied to temporal logics.
The Javert miner~\citep{Gabel2008Javert} 
finds temporal properties using small generic patterns that are 
composed soundly into larger specifications, and establishes
this form of mining is NP-Complete. 
~\cite{Reger2013Pattern} extends mining with patterns to parametric 
temporal specifications expressed as quantified event automata. 
The Texada General LTL Specifications Miner~\citep{Lemieux2015Texada}
accepts user-defined LTL templates and studies traces 
to produce instantiations of the LTL formula.

While first demonstrated for use in software engineering, 
the generality of Texada allows it to be easily extensible.
Texada supports
arbitrary templates and is helpfully maintained as an open source tool.
For this reason, Texada is used within
the Undine framework, where Texada's template enumerator 
is modified to incorporate
a labelling system
on trace events.

\section{Automatic Software Security} 

In the software domain a number of technologies, including specification
mining, have been applied to develop security specifications automatically.

AutoISES~\citep{Tan2008AutoISES} uses security check rules and static 
analysis of source code to 
automatically generate security specifications of operating systems.
Juxta~\citep{Min2015Juxta} also uses static analysis, in this case applied
to the Linux file system and extensible to 
web browsers, network protocols, and other forms of software with multiple
implementations.
ClearView~\citep{Perkins2009ClearView} is a security autopatcher for
Windows that works on binaries rather than source code with a 
specification mining component that studies normal behavior to devise error
detectors.
~\cite{Yamaguchi2011Vulnerability} datamines API usage patterns
which, when provided with a known vulnerability, enable automatic 
extrapolation of vulnerabilities over libraries.

\section{Automatic Hardware Functional Correctness} 

Specification extraction has been applied to hardware, usually run
in simulation, with different considerations made to handle hardware
complexity.

IODINE~\citep{Hangal2005Iodine} 
applied automatic specification generation to hardware by looking for
instances of known patterns, such as one-hot encoded
signals and req-ack. Reflecting the origins of the technique in the
software domain, IODINE continues the research directions of both
 DIDUCE~\citep{Hangal2002Diduce} and
Daikon~\citep{Ernst2007Daikon} from the software domain, applying
optimizations from both online and offline software mining to generate results
over hardware in an offline context.

Rather than rely on patterns,
~\cite{Chang2010Automatic} use sequential data mining of simulation traces
for automatic exploit detection specifically around potentially 
malicious inputs.
~\cite{Liu2011Automatic} perform mining at a higher abstraction level by
studying Transaction Level Model (TLM) simulated traces.

~\cite{Mandouh2012Automatic} also use hardware specific patterns,
but use static analysis to generate them before moving to a dynamic
stage. GoldMine~\citep{Hertz2013GoldMine} mines traces for specific
patterns, and also includes manual efforts in a 
 stage where designers rank assertions.

More recent work
has focused on mining temporal properties from execution
traces.
Similar to techniques in software with temporal specification,
~\cite{Li2010Scalable} 
use predefined patterns and pattern chaining 
to mine temporal properties from traces.
~\cite{Liu2012Word} mine over traces specified 
at word level rather than bit level, and shows that 
experimentally
this results in higher expressiveness, including on RISC CPU designs.
Behavioral models~\citep{Danese2015Automatic}
and power state machines of SoC designs~\citep{Danese2016Automatic} both may also
provide useful abstractions to study hardware. 
The A--TEAM miner~\citep{Danese2017ATEAM} is able to mine designs
given LTL templates.

Most recently, ~\cite{Rawat2020Hyperminer} developed algorithms to mine
hyperproperties expressed in HyperLTL using trace fuzzing. Their framework,
Hyperminer, finds
useful hyperproperties, including noninterference, an information flow property,
 over a small SoC design.
In this respect, Hyperminer is similar to Isadora which produces information
flow properties of small SoCs and other designs.
By way of contrast, Isadora additionally
describes flow conditions, or conditional interference patterns, which capture
common security goals where design elements must interact but only with
certain privileges, permissions, or under some other design condition.

\section{Manual Hardware Security}

The first security properties developed for hardware designs were manually
crafted. 
Security Checkers~\citep{Bilzor2011SecurityCheckers} uses manually defined
hardware design language (HDL) assertions to 
generate hardware runtime monitors.
SPECS~\citep{Hicks2015SPECS} also uses manual assertions but 
works on only security critical state at the ISA level to reduce overhead
and run on a full RISC processor.
~\cite{Brown2017Thesis} reviewed the Intel Software
Developers Manuals~\citep{Intel2020SDM} to manually define security
critical properties at ISA level for x86.
Recent case studies~\citep{Dessouky2019HardFails} have revealed the types of
properties needed to find exploitable bugs in the design of a RISC-based
system-on-chip.

\section{Automatic Hardware Security}

SCIFinder~\citep{Zhang2017SCIFinder} semi-automatically generates
security critical properties by using machine learning to label 
generated invariants based on similarity to known bugs.
Transys~\citep{Zhang2020Transys} is able to automatically generate
security properties for a target design by translating known properties
from some other design to analogous properties on the original design,
but does require some initial set of known security properties.

\section{Contextualizing this Thesis}

This thesis considers security specification mining with
intent to support security validation of hardware designs. 
To do so, each
miner takes lessons from 
across each dichotomy, and adapts
them to work specifically on hardware security. For example, the Undine
miner, which creates linear temporal logic (LTL) specifications, 
uses the Texada LTL Specifications Miner, 
and the Astarte miner, which discovers invariants
using control signals, and the Isadora miner,
which generates predicates that condition information flows, 
use the Daikon Dynamic Invariant Detector.
However, Texada and Daikon are 
tools for automating software functional correctness.
To use these tools for hardware security
requires synthesizing insights from both applying specification mining to hardware,
such as the patterns seen in IODINE and other miners, and from distinguishing
security properties from functional properties, such as in SPECS and SCIFinder. 
In isolation, 
IODINE, A-TEAM, and other miners do not target security specifications, and
SPECS and SCIFinder do not automatically
generate information flow or LTL specifications or specifications for CISC designs.
To automatically generate security specifications of hardware designs that
cover known and potential future attacks, the work presented in thesis
develops hardware security specification mining 
over different designs and logics of specification.

\subsection{Astarte}

In the case of Astarte, which targets the Instruction Set Architecture
level, existing tools including the Daikon Dynamic Invariant Detector
readily produce trace invariants given some form of trace data.
However, Daikon does not innately generate properties of the form $A \implies B$.
Daikon allows users to specify possible predicates 
for the antecedent $A$, and will then generate implication properties which
specify some consequent $B$.
Many manual efforts to describe secure behavior of processor
designs frequently 
contain implications
~\citep{Bilzor2011SecurityCheckers,Hicks2015SPECS,Brown2017Thesis,Bilzor2012psl2hdl}.
One approach may be to specify every possible predicate $A$, but this approach is
exceedingly costly over x86, the studied design, which contains
too many possible antecedents to consider exhaustively.
Further, this approach would produce many properties with no notion
of their relevance to preventing future attacks.

The Astarte framework approaches the challenge of dynamic invariant
detection over x86 using the notion of security critical control signals.
Astarte iteratively performs invariant detection to restrict
the space of candidate predicates and then uses these predicates
to discover a property set
expressive enough to contain manually developed security properties, 
including implications, yet excluding many redundant 
or uninteresting properties.

\subsection{Undine}

Undine approaches the challenge of generating
LTL properties over up to three propositional variables 
to include register, subfield, bit, and delta values and equality between
registers over gigabyte trace sets.
Existing hardware miners could address subsets of these events or 
templates of up to two variables, and 
Texada could mine properties of this form generally but faced challenges with regard 
to complexity.
Additionally, Undine targets output specifications that specifically
prevent attacks.
When approaching the trace sets needed to converge on a steady state
of properties for open source RISC designs, Texada struggled to produce
properties in under four hours due to the number of unique variables
in RISC traces, which were approximately three times greater than the 
maximum of 1000 unique variables used to evaluate Texada by ~\cite{Lemieux2015Texada}.
This is explored in depth in Section~\ref{sec:vsTexada}, where the Undine
workflow is evaluated with unmodified and partially modified Texada instances
to contextualize Undine with prior art on different templates.

Undine can also be considered in the context of temporal hardware miners.
SAM~\citep{Li2010Scalable} extracts expressive properties using many scalability
optimizations, including submodule and subtrace decomposition. 
SAM uses delta event traces
which only capture register values, and extension
to consider equality between registers would be nontrivial.
A--TEAM extracts properties of the form 
$\mathbf{G}(\theta \implies \psi)$ by combining coverage analysis with data mining.
By contrast, Undine is able to consider templates over three variables 
yet innately contains no coverage analysis.
With respect to coverage for Undine, 
trace volume needed to converge to a steady set of properties was studied
and is discussed in Section~\ref{sec:traceslength}.

\subsection{Isadora}

In the case of Isadora, existing specification miners with the exception
of Hyperminer were unable to produce information flow properties.
Hyperminer produces noninterference properties over pairs of 
design elements using template enumeration (the technique used in Undine).
An Isadora information flow specification contains two parts.
The first part is
the ``Never-Flow Pairs'' of Section \ref{sec:propertymining} which
are equivalently expressive to properties produced by Hyperminer using the
noninterference template.
The second part is 
the ``Flow Conditions'' which are equivalently expressive to declassification, 
a form of information flow property distinct from noninterference (and
dynamic determinism, the other demonstrated Hyperminer template).

It is likely the case that Hyperminer template enumeration could
be extended through further efforts to
target declassification properties. This would be logically similar
to property development in Isadora, which at a 
high level used IFT
to determine interference patterns and from there composed declassification
properties. This process and more comparisons to Hyperminer are
discussed in Chapter~\ref{chap:Isadora}.
\chapter{Astarte: Mining CISC Architectures}
\label{chap:Astarte}

\section{Introduction}

Many existing property specification tools were developed for, 
and are applicable to, open-source programs with respect to functional 
correctness. Even when applied to hardware, tools often only work with
open source designs, such as OR1200 or RISC-V.
Considering the x86 instruction set architecture (ISA) introduces three main 
challenges. 
First, the size of the ISA makes even a semi-manual approach prohibitive; 
Second, x86 is closed-source and existing approaches to mining security 
specifications relied on access to both the source code and the 
developers' repositories, bugtracker databases, and email 
forums~\citep{deutschbein2018undine}; 
Third, compared to today's RISC architectures, x86 offers a richer
landscape of security features and privilege modes, increasing the number and
complexity of the associated security properties.

To overcome these challenges, I developed Astarte, a fully automatic security 
specification miner for x86. 
On x86, the key challenge with mining security critical properties
is automatically identifying those properties that are relevant
for security, that if violated would leave the processor vulnerable to
attack. 
In general, there is no fixed line separating functional properties
from security properties. The environment in which a processor
operates and the attacker's motivation and capabilities may move some properties
across the security-critical boundary in either direction.

Prior work tackled this problem by analyzing existing design bugs and manually 
sorting them as exploitable or not exploitable~\citep{Zhang2017SCIFinder}.
However, this approach is labor intensive and does
not easily scale to x86. 
Further, this approach requires 
knowledge of and access to the details of known design bugs culled from
developers' archives, code repositories, and bugtacker databases, which are
not available for the closed-source x86 designs.

Astarte uses a different approach.
By mining properties that are conditioned on the state of the various control 
signals that govern security-critical behavior of the processor, Astarte
need not rely on inaccessible documentation. 
The corresponding properties are by definition important for the correct and 
secure behavior of the processor, which in turn is important for the correct 
implementation of the security primitives that operating systems (OSs) and software 
rely on. 
In this respect, Astarte is inspired by prior, manual
efforts~\citep{Bilzor2011SecurityCheckers,Hicks2015SPECS,Brown2017Thesis,Bilzor2012psl2hdl}.

Astarte handles the complexity of the x86 ISA by independently considering the
space of properties for each instruction preconditioned on the value of a single
security-relevant control signal. 
In other words, the mining partitions the specification generation problem with 
respect to each control signal. 
It is perhaps counter-intuitive that this approach works; it would seem necessary
to consider all possible combinations of all security-relevant signals for every 
instruction in order to produce meaningful security properties. 
Yet, compared to prior manual efforts and to known bugs in shipped x86 products, 
the specification output of Astarte independently produces valuable properties
using this technique.

The Astarte framework relies on two existing tools, the QEMU 
emulator~\citep{Bellard2005QEMU}, which creates traces over x86, and the Daikon 
Dynamic Invariant Detector~\citep{Ernst2007Daikon}, a popular tool for mining 
specifications of programs. 
Using the QEMU debug interface, I generated ISA level traces of x86 running
four distinct operating systems and bare metal programs.
Astarte incorporates a custom front end for Daikon to interpret the QEMU debug
logs as trace data and mine invariants over these emulated runs.
Daikon, a major research and software engineering effort extending
for decades, offers modular interfaces for this custom front end
and a powerful internal inference engine. Together these
provide a strong foundation for scalability to studying the x86 
architecture.

When run on emulated traces, Astarte produces roughly 1300 properties.
I evaluated these properties against 29 security properties manually
discovered by~\cite{Brown2017Thesis}.
Of the 29 identified security properties, Astarte generates 23, and the 
remaining 6 require invariants over processor state unimplemented in QEMU.
Astarte also generates properties that, if followed, would prevent 
two bugs in x86 documented in public domain, Memory 
Sinkhole~\citep{Domas2015Sinkhole} and SYSRET privilege 
escalation~\citep{Xen2012SYSRET}. 

By performing trace generation over multiple operating systems and bare metal
execution, I show Astarte can also differentiate properties enforced by the 
processor from those that must be enforced by the operating system.
This analysis also provides insight into properties
that are not specified, but that operating systems have come to rely on. 

This chapter presents a security specification miner for closed-source, x86
architectures and its evaluation for the Intel x86 (Ivy Bridge) processor. 
Astarte demonstrates:
\begin{itemize}
  \item partitioning specification ming using
    security-relevant control signals; 
  \item automatically identifying the control signals of interest;
  \item differentiating processor-level properties and operating
    system-level properties; and
  \item identifying de facto security-critical properties upon which operating
    systems rely.
\end{itemize}
 
\section{Properties}

The Astarte framework generates properties written over instruction
set architecture (ISA) level expressions
conditioned on current instructions and automatically discovered 
control signals.
This section discusses the structure of these properties
and describes which properties are considered to be security properties
for this framework.

\subsection{Example Properties}

Astarte properties describe the constraints and behavior of ISA level state. 
The grammar is described in Figure~\ref{fig:astgrammar},
where $\mathtt{insn}$ represents the predicate for some
specific instruction and 
$\mathtt{sig}$ represents the valuation of a signal. 
Signals here can represent software visible registers
or individual bits within some specific registers. The
process of determining individual bits for consideration 
is detailed in Section ~\ref{sec:idsignals}.

Each property produced by Astarte describes the processor state
as an implication.
The antecedent may be a predicate specifying that the property describes
processor state
during the execution of a particular instruction,
specifying that some expression over signals holds, or specifying both.
For brevity, these antecedent predicates 
will be referred to as preconditions.
The consequent specifies an expression over signals.
Expressions over signals, in either antecedents or consequents, may be:
\begin{itemize}
\item equalities between the values of two signals,
\item equalities between the value of a signal and a constant, 
\item a signal's value falling within a set of up to three constants, or 
\item equality between the current value of a signal and the previous
value of the same signal, which is specified using $\mathit{orig}()$.
\end{itemize}

\begin{figure}
  \centering
\begin{align*}
  \phi \doteq&~ \mathtt{insn} \land e_1 \implies e_2 ~|~ e_1 \implies e_2 ~|~ \mathtt{insn} \implies e\\
  e \doteq&~ \mathtt{sig_1} = \mathtt{sig_2} ~|~ \mathtt{sig} = n  ~|~  \mathtt{sig} \in \{n_1,n_2,n_3\} ~|~ \mathtt{sig} = \mathit{orig}(\mathtt{sig})
  \\
\end{align*}
\caption{The grammar of Astarte properties}
\label{fig:astgrammar}
\end{figure}

\begin{figure}[h]
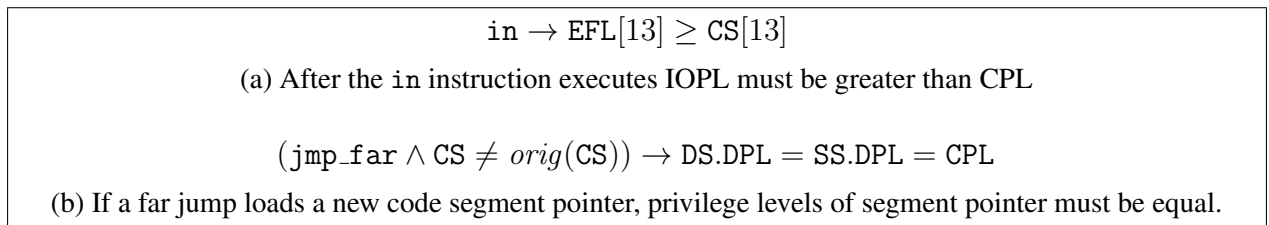

  \centering
  \begin{subfigure}{\columnwidth}
    \centering
  $\mathtt{in} \rightarrow \mathtt{EFL}[13] \geq \mathtt{CS}[13]$
    \caption{After the $\mathtt{in}$ instruction executes IOPL must be greater than CPL}
    \label{fig:examplepropertyA}
  \end{subfigure}\\\vspace{15pt}
  \begin{subfigure}{\columnwidth}
    \centering

  $(\mathtt{jmp\_far} \land \mathtt{CS}\neq \mathit{orig}(\mathtt{CS}))\rightarrow \mathtt{DS.DPL} =\mathtt{SS.DPL} = \mathtt{CPL}$
    \caption{If a far jump loads a new code segment pointer, privilege levels
    of segment pointer must be equal.}
    \label{fig:examplepropertyB}
  \end{subfigure}
  \caption{Example properties}
\end{figure}

For example, the property in Figure~\ref{fig:examplepropertyA} states that when the $\mathtt{in}$ instruction executes,
the I/O privilege level (IOPL, as given by bit 13 of the EFLAGS register) must be greater than or
equal to the current privilege level (CPL, as given by bit 13 of the Code
Segment Pointer). In this case, this is an property found in the 
context of a single instruction.

A property may refer to the value held by a bit or register both before
and after the relevant
instruction executes by using an $\mathit{orig}()$ expression 
to describe the original value before
execution.
 For example, the property in
Figure~\ref{fig:examplepropertyB} refers to the state of the
Code Segment Pointer (CS) before and after execution of the $\mathtt{jmp\_far}$
instruction. 
It states that if a new CS is loaded by a far jump, then the privilege levels of
the other current segment pointers, the Stack Segment Pointer (SS) and Data Segment 
Pointer (DS),
 must be equal. Here DPL denotes `descriptor privilege level' 
and is the privilege of a given segment pointer; the CPL is simply the 
DPL for the CS.
In this case, this is a property found using a precondition 
that the CS value after instruction executes must differ from the CS 
prior to execution.
Of note, while near and far jumps are not differentiated by name in emulation,
the Astarte trace encoding stage inspects opcodes to differentiate near and
far cases, and appends the $\mathtt{\_far}$ suffix at this time.

\floatstyle{boxed}
\restylefloat{figure}

\floatstyle{plain}
\restylefloat{figure}

Given the constraints of generating traces over a closed source design,
the example properties here describe values as reported
through emulation debug interfaces. 
Astarte assumes the correctness of emulation. 
Prior work, such as Fast PokeEMU ~\citep{yan2019FastPokeEMU},
addresses directly the correctness of CPU emulation.

\subsection{Control Signals for Preconditions}

Key to the working definition of security property for Astarte is the notion
of a control signal. Within Astarte, control signals are registers or bits
which, when included in the trace generation stage within preconditions,
produce output properties distinct from the set of output properties
discovered when not using preconditions. The precise preconditions are
described in detail in Section ~\ref{sec:precons}.

Should some signal, when used as a precondition for
extracting properties, produce a distinct property set,
the signal is then associated with some distinct behavior of the processor,
such as that the signal's value may only be changed under certain conditions, that 
certain instructions may operate in a particular way while the signal is set,
or that, when set, values of other design elements may or may not be updated. 
In brief,
these signals control the allowable behavior of a design.

\subsection{Security Properties for Astarte}

Astarte is intended to discover properties that describe the intended 
secure behavior
of processor designs. Astarte should capture some notion
of software's expectations of hardware, such as by enforcing control bits
exercised by operating systems including the IOPL and CPL.
Astarte should also provide coverage
of some security properties discovered through manual review of documentation,
and discover properties that, if followed, would prevent processors from
being vulnerable to documented historical attacks.
Ideally, Astarte would not be limited to just these cases and may additionally
discover unanticipated properties that may aid hardware designers in
ensuring secure behavior of the processor.
Using control signals within property
preconditions may address each of these goals.

For this reason, within the context of the Astarte framework,
the working definition of a security property is a property over the
design describing allowable designs states either when a control
signal is fixed or when a control signal value 
is updated. Some of the limitations of this working definition
are explored in the context of the results in
Section ~\ref{sec:functionalprops}.

\subsection{Properties in Implementation}

\floatstyle{boxed}
\restylefloat{figure}

\begin{figure}[h]
    \begin{verbatim}
"..jmp_far():::EXIT;condition="not(CS==orig(CS))""
['1', 'A20', 'CR0_0', 'CR0_1', 'CR0_18', 'CR0_3', 'CR0_5', 
'CR4_6', 'CR4_9', 'D_CPL', 'D_CR0_0', 'D_CR0_1', 'D_CR0_18', 
'D_CR0_2', 'D_CR0_20', 'D_CR0_3', 'D_CR0_5', 'D_CR0_7', 
'D_CR4_11', 'D_CR4_12', 'D_CR4_6', 'D_CR4_8', 'D_CR4_9', 
'D_EFL_1', 'D_EFL_11', 'D_EFL_13', 'D_EFL_2', 'D_EFL_4', 
'D_EFL_6', 'D_EFL_8', 'D_EFL_9', 'D_SMM', 'EFL_1', 'EFL_9', 
'orig(A20)', 'orig(CR0_0)', 'orig(CR0_1)', 'orig(CR0_18)', 
'orig(CR0_3)', 'orig(CR0_5)', 'orig(CR4_6)', 'orig(CR4_9)', 
'orig(D_CR0_0)', 'orig(D_CR0_1)', 'orig(D_CR0_18)', 
'orig(D_CR0_2)', 'orig(D_CR0_20)', 'orig(D_CR0_3)', 
'orig(D_CR0_5)', 'orig(D_CR0_7)', 'orig(D_CR4_11)', 
'orig(D_CR4_12)', 'orig(D_CR4_6)', 'orig(D_CR4_8)', 
'orig(D_CR4_9)', 'orig(D_EFL_1)', 'orig(D_EFL_11)', 
'orig(D_EFL_13)', 'orig(D_EFL_2)', 'orig(D_EFL_4)', 
'orig(D_EFL_6)', 'orig(D_EFL_8)', 'orig(D_EFL_9)', 
'orig(D_SMM)', 'orig(EFL_1)', 'orig(EFL_9)']
['0', 'CR0_2', 'CR0_20', 'CR0_7', 'CR4_11', 'CR4_12', 'CR4_8', 
'DR1', 'EFER', 'EFL_11', 'EFL_13', 'EFL_4', 
'EFL_8', 'FS_DPL', 'GS_DPL', 'HLT', 'II', 'LDT_DPL', 'SMM', 
'orig(CR0_2)', 'orig(CR0_20)', 'orig(CR0_7)', 'orig(CR4_11)', 
'orig(CR4_12)', 'orig(CR4_8)', 'orig(EFL_11)', 'orig(EFL_13)', 
'orig(EFL_4)', 'orig(EFL_8)', 'orig(GS_DPL)', 'orig(HLT)', 
'orig(II)', 'orig(LDT_DPL)', 'orig(SMM)']
    \end{verbatim}
    \caption{All signals equal to zero or one
    during far jumps across code segments.}
    \label{fig:zeroone}
\end{figure}
\floatstyle{plain}
\restylefloat{figure}

To consider the output properties of Astarte, Figures~\ref{fig:zeroone}
and~\ref{fig:dpls} shows the invariant with the precondition of a far jumps
that loads a new code segment pointer. This is the condition referenced
by the example property in Figure~\ref{fig:examplepropertyB}. 

In these figures, the first line gives the preconditions, and each successive
line (which may be wrapped) contains tuples of values or signals
which are found to be equal within the trace set when this condition holds.
The most frequent cases are comparisons to zero or one, which
are broken out specifically in Figure~\ref{fig:zeroone}.
The size of this
output is greatly inflated by considering individual bit values and
their delta values (with the `D\_' prefix)
within many registers and considering each of these 
both before and after the instruction executes.
Delta values are not noted in the grammar and are not
 necessary to the Isadora workflow, but deriving
 these values
offered niceties in implementation, and they are included here for
completeness.
The delta values for bits are incremented by one so that they do not take on a 
negative value which is used for error checking within the workflow,
so `$\mathtt{orig(D\_CR0\_0)==1}$' denotes that the 0th bit of the CR0
register was unchanged from the previous clock cycle before the far 
jump executes. This bit encodes whether the design is in protected mode,
and means the trace set contains no case of a change to
protected mode status immediately prior to a
far jump that loads a new CS pointer.

\floatstyle{boxed}
\restylefloat{figure}

\begin{figure}[h]
    \begin{verbatim}
"..jmp_far():::EXIT;condition="not(CS==orig(CS))""
...
['EFL_2', 'orig(EFL_2)']
['CPL', 'CS_DPL', 'DS_DPL', 'ES_DPL', 'SS_DPL']
['4294905840L', 'DR6', 'orig(DR6)']
['"0080c1100a800000207300008900"', 'TR', 'orig(TR)']
['1024', 'DR7', 'orig(DR7)']
['CR2', 'orig(CR2)']
['CR3', 'orig(CR3)']
['IDT', 'orig(IDT)']
['EFL_11', 'HLT', 'orig(EFL_11)', 'orig(II)']
['A20', 'D_EFL_11']
    \end{verbatim}
    \caption{All remaining signal equalities
    during far jumps across code segments.}
    \label{fig:dpls}
\end{figure}
\floatstyle{plain}
\restylefloat{figure}

Beyond signals found to be equal to zero or one,
the example property in Figure~\ref{fig:examplepropertyB} appears
within the output.
The second line in Figure~\ref{fig:dpls} shows that, over
properly encoded traces and under the relevant
preconditions, Astarte may detect an equality between the 
descriptor privilege levels (DPLs) of
the code segment, stack segment, and data segment pointers
after a far jump executes. 
Of note, CPL and CS\_DPL are both listed
because both the emulator and Astarte derive the value independently.
QEMU only derives the CPL and Astarte derives the DPL for all
segment pointers; the two are always equal because they 
refer to the same bits in CS.

This output additionally specifies the values or equalities
of other registers.
The Extra Segment Pointer (ES) is also found to have a 
descriptor privilege
level equal to that of the CS, SS, and DS.
DR7, a debug register, is found to always hold the value 1024.
DR7 is frequently set to 1024, and it fact is set
to 1024 in all cases in operating system traces.
CR3, the Page Fault Linear Address for recovery
after page faults, is also held constant across code 
segment changes. CR3 is updated within the trace set but never
during a far jump.

Astarte's output properties can appear large and complex due to the 
internal implementation details and scope of design state,
but they conveniently define processor conditions within a single
instruction and precondition and simply describe the software
visible state as emulated and logged at these points.
\section{Methodology}

\begin{figure}
\centering
\includegraphics[width=\columnwidth]{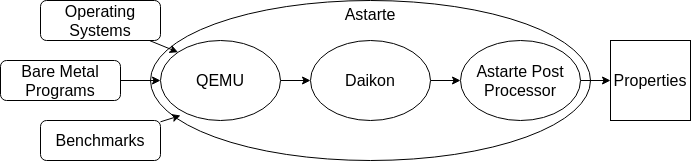}
\caption{An overview of the Astarte workflow}
\label{fig:astarteoverview}
\end{figure}

Astarte works in three phases: trace generation, property mining, and
post-processing.  Figure~\ref{fig:astarteoverview} provides an overview of the Astarte
workflow. In the first phase I generate traces of execution of the
processor. Without access to the source code of the processor design, I can not
use a simulator to generate traces of processor execution as in prior work. 
Instead I use QEMU,
an x86 emulator, to emulate processor execution. QEMU translates blocks of code
at a time, and as such produces traces of basic blocks. The miner requires traces of
individual instructions, so Astarte extends the generated traces so that each
event in the trace represents a single
instruction. This extension is sound with respect to the generated properties. In
keeping with prior art, Astarte tracks processor state that is
visible to software; the final security properties are written over this
software-visible state. I emulate the processor loading and running four
different operating systems as well as running software on the bare (emulated)
metal.

In the second phase Astarte mines the traces of execution looking for security
properties. I build the miner on top of the Daikon invariant generation
tool~\citep{Ernst2007Daikon}. The closed-source nature of x86 processors precludes
using known, exploitable design bugs to differentiate security-critical
properties from functional properties as was done by~\cite{Zhang2017SCIFinder} when
targeting RISC processors. Furthermore, the complexity of the many x86
protection modes and their associated control flags overwhelms the miner. In an
initial experiment letting a naive miner run for 7 days, 
hundreds of millions of invariants were generated, with only a fraction 
representing
useful security properties. 
To address this, Astarte partitions
the state space of the processor on each of a small handful of security-critical
control signals, and generates invariants within each partition. 
It is perhaps counter-intuitive that this approach works; it would seem necessary
to consider all possible combinations of all security-relevant signals for every 
instruction in order to produce meaningful security properties. 
Yet, compared to prior manual efforts and to some known bugs in shipped x86 products, 
the specification output of Astarte independently produces valuable properties
using this technique.

In the third phase, post-processing, Astarte combines like invariants, integrates
results across multiple runs of the miner (e.g., using traces generated from
different operating systems), and simplifies expressions. The result is a
manageable set of security properties.

\subsection{Trace Generation}
To generate traces of execution I use QEMU, a
full system, open source machine emulator~\citep{Bellard2005QEMU}. 
Running a processor in emulation
allows us visibility into the processor's state. 
QEMU-emulated x86 CPUs can boot
operating systems, run user-level applications, and output log data about the
state of the CPU as software executes. The log data forms the basis of the
execution traces over which security properties are mined.

QEMU dynamically translates machine instructions from the target architecture
(in this case, x86) to the host architecture.
To ease portability QEMU
translates first to an intermediate language and then to the host
instruction set. To improve performance
QEMU translates a block of machine instructions at a time, rather than
translating line by line.
While this feature may be deactivated, doing so prevented emulation from being
able to boot most modern OSs in under a day and greatly increases trace size.

A QEMU translation block (TB) is akin to a basic block~\citep{Aho2006Compilers}. 
It is a
sequence of instructions with a single entry point---the first instruction in the
TB---and a single exit point---the last instruction in the TB. A TB ends at any
instruction that modifies the program counter, such as syscall, sysenter, or
jmp, or at a page boundary.  

The translated TBs can be cached and reused, reducing translation time. However,
translating a block of code at a time obscures CPU state at instruction
boundaries. In other words, QEMU maintains consistent
target CPU state at TB boundaries rather than at instruction boundaries.

From the QEMU execution logs, Astarte may extract \emph{events} corresponding 
to the execution of a single TB. An
event shows the sequence of instructions that make up the TB and the CPU
state after the TB executes:
\begin{align*}
  \langle\mathrm{instruct}_1,\mathrm{instruct}_2,\ldots,\mathrm{instruct}_n\rangle(r_0, r_1,
  \ldots, r_m)
\end{align*}

In the above, ($r_0, \ldots, r_m$) represents the state of the $m$ ISA-level registers
after the TB executes. A \emph{trace} of events gives the CPU
state at every TB boundary. The first
event in a QEMU trace is always the single instruction \texttt{ljmpw}, which jumps to the code
entry point. A trace of TBs might look like this:
\begin{align*}
  \langle\mathtt{ljmpw}\rangle&(r_0^0,r_1^0,\ldots,r_m^0)\\
  \langle\mathrm{instruct}_1,
  \mathrm{instruct}_2,\ldots,\mathrm{instruct}_n\rangle&(r_0^n,r_1^n,\ldots,r_m^n)\\
  \ldots&\\
  \langle\mathrm{instruct}_1,\mathrm{instruct}_2,\ldots,
  \mathrm{instruct}_j\rangle&(r_0^*,r_1^*,\ldots,r_m^*)
\end{align*}
The CPU state logged at the end of the first TB gives us the CPU state before
the second TB executes. 
Consider this compared to a trace specifying each instruction. The single logged event
$\langle\mathrm{instruct}_1,\mathrm{instruct}_2,\ldots,\mathrm{instruct}_n\rangle(r_0^n,
r_1^n, \ldots, r_m^n)$, would correspond to an extended trace of events:
\begin{align}
  \langle\mathrm{instruct_1}\rangle&(r_0^1,r_1^1,\ldots,r_m^1)\\
  \langle\mathrm{instruct_2}\rangle&(r_0^2,r_1^2,\ldots,r_m^2)\label{eq:line2}\\
  \ldots\nonumber\\
  \langle\mathrm{instruct_n}\rangle&(r_0^n,r_1^n,\ldots,r_m^n)
\end{align}

Producing the extended trace of events would require the emulator translate
code line-by-line. But, the emulator still needs to be fast enough to boot
operating systems and run application-level code. 
I take an intermediate approach for trace generation: 
I use QEMU translation blocks and build a lightweight extension to
generate partial per-instruction events. For every TB in a trace, the event 
generator creates a
new sequence of events, one event for each instruction in the TB. Each event
lists the instruction executed and partial information about the CPU state. Any
software-visible register that can be modified by the instruction is marked as
invalid, and all other registers retain their value from the previous event. The
generated event corresponding to the last instruction in the TB has the full CPU
state as given by the original QEMU event. Continuing with the above extended
trace of events, and considering the second event at line (\ref{eq:line2}) in
the trace, $\forall i, 0\leq i\leq m$ either $r_i^2=r_i^1$ or $r_i^2=\texttt{invalid}$.

The
event generator errs on the side of soundness: if it is possible for an
instruction to change an aspect of CPU state, the generator assumes it does. I used the
Intel 64 and IA-32 Architectures Software Developer Manuals~\citep{Intel2020SDM} as reference when building
the event generator. 

\subsection{Property Mining}
\label{sec:propertymining}

Astarte uses Daikon~\citep{Ernst2007Daikon} as the base for
property mining. I
build a custom front-end that reads in the extended traces of events produced in the
first phase, and outputs a trace of observations suitable for Daikon.

Daikon was
developed for use with software programs: it looks for invariants over state
variables for each point in a program. This front-end treats x86 instructions as
program points; Daikon therefore will find invariants over ISA variables for
each x86 instruction.

Daikon can handle individual program modules with relatively few
program points and few program variables, it is not intended for analysis of
entire programs. The amount of ISA state and the number of
instructions in x86 is too large for Daikon to handle. The amount of trace data required to
achieve coverage of a single instruction, and the size of the state over
which to find invariant patterns for a single instruction overwhelm Daikon.

To mitigate the complexity, for each instruction Astarte partitions the space of
properties on individual control signals. 

\subsubsection{Partitioning on Control Signals}
\label{sec:precons}
For each instruction, Astarte separately considers the space of invariants
over ISA state for that instruction, preconditioned on a single control bit. The
key insight is that if Astarte chooses the control bits wisely, the partitioning not only
mitigates performance and complexity issues with Daikon, it also produces sets
of properties that are critical to security, and can then classify the
properties by their precondition. The properties that make up each class provide
some insight into the modes and behaviors of the processor governed by the
preconditioning control signal.

For each control signal, how Astarte partitions the space of invariants for a single
instruction depends on the control signal. For a one-bit signal Astarte creates four partitions, one for
each combination of signal values before and after the instruction executes. For
example, with the $\mathtt{IOPL}$ flag and $\mathtt{addl}$ instruction, Table~\ref{tab:expartition} shows the four partitions of the space of
invariants. Each row of the table represents one of the four possible
antecedents of a property. The four antecedents represented in the table
completely partition the space. For signals longer than one bit Astarte divides the
space of invariants into two partitions for each instruction: $\mathtt{instruct} \wedge \mathit{orig}(\mathtt{reg}) = \mathtt{reg}$
and $\mathtt{instruct} \wedge \mathit{orig}(\mathtt{reg}) \neq \mathtt{reg}$.

The set of properties produced for a particular preconditioning signal tell us
something about the behavior governed by that signal. For example, providing $\mathtt{CPL} \neq \mathrm{orig}(\mathtt{CPL})$ as a
  precondition will mine properties related to how the current privilege level
  (CPL) of the processor is elevated and lowered.

  \begin{table}[t]
    \centering
    \begin{tabular}{ccc}
      \toprule
      $\mathtt{addl}\quad \wedge $ & $\mathit{orig}(\mathtt{IOPL})=0 \quad\wedge$ & $\mathtt{IOPL}=0$\\
      $\mathtt{addl} \quad\wedge $ & $\mathit{orig}(\mathtt{IOPL})=0 \quad\wedge$ & $\mathtt{IOPL}=1$\\
      $\mathtt{addl} \quad\wedge $ & $\mathit{orig}(\mathtt{IOPL})=1 \quad\wedge$ & $\mathtt{IOPL}=0$\\
      $\mathtt{addl} \quad\wedge $ & $\mathit{orig}(\mathtt{IOPL})=1 \quad\wedge$ & $\mathtt{IOPL}=1$\\
      \midrule
    \end{tabular}
    \caption{Astarte partitions shown on the $\mathtt{IOPL}$ signal for instruction $\mathtt{addl}$}
    \label{tab:expartition}
      \end{table}
  
\subsubsection{Identifying Control Signals}
\label{sec:idsignals}

The first step is to choose which control signals to use as preconditions. I
manually organized the x86 ISA state by category and then let Astarte find the
meaningful signals within a category. 
The registers and their categories are given in Table~\ref{tab:categories}.

\begin{table}[t]
  \centering
  \begin{tabular}{ll}
\toprule
Category & Registers\\
\midrule
General Purpose Registers & EAX, EBX, ECX, EDX\\
 Interrupt Pointer & EIP\\
 Control Registers & EFL, CR0, CR2, CR3, CR4, EFER\\
 Individual Bitflags & II, A20, SMM, HLT, CPL\\
 Current Segment Pointers & CS, SS, DS\\
 Special Segments Pointers & ES,  FS, GS, LDT, TR\\
 Descriptor Tables & GDT, IDT\\
 Debug Registers & DR0, DR1, DR2, DR3, DR6, DR7\\
 Command Control & CCS, CCD, CC0\\
 \midrule
  \end{tabular}
  \caption{Categorization of QEMU logged registers}
  \label{tab:categories}
\end{table}

Of these, Astarte focuses on the three control encoding categories: 
the Control Registers, the Individual Bitflags, and the Current Segment Pointers. 
These design elements either themselves control or
contain fields that control security critical state, such as privilege levels
and location of page tables. I chose these categories based on knowledge of
the x86 ISA. Initially, Astarte used only the Control Registers and Bitflags, but
initial evaluation led me to add the Current Segments. It is possible other
categories may also yield interesting properties. 
Fortunately, each control signal is analyzed independently of the others, so
additional categories of ISA state can be analyzed without incurring a
combinational explosion in performance cost. (In Section~\ref{sec:EvalOptim}
I discuss the cost.)

During the signal-finding phase Astarte unpacks registers to consider one- and 
two-bit fields separately.
Mining can be used to reveal many
fields that keep a constant value which are discarded as unused fields. 
To remove these from the considered fields, Astarte collapses all x86 instructions into a
single pseudo-instruction and runs the property miner on this modified trace. 
Any found properties of the form $\mathtt{reg}=N$ are an indication that for all
instructions $\mathtt{reg}$ has the constant value $N$ and $\mathtt{reg}$is 
therefore unused. Astarte discards
these flags from further consideration. At the end of this phase Astarte is left
with 24 signals of interest that either single bits, two-bit fields, or registers.

\subsection{Postprocessing}

The Daikon miner produces tens of millions of properties. In post-processing Astarte
removes invalid properties, removes redundant properties, and combines similar
properties into a format that is easier to read.

\subsubsection{Intersection Across Trace Sets}
Astarte runs the Daikon miner separately for each set of traces
representing separate operating system boots and bare-metal execution. In the
first step of post-processing, properties from different traces are combined by taking
the intersection of all sets with shared elements within a precondition.
This ensures that no property that is invalidated by any one trace
is considered as an output property.  It also generalizes properties
to the implementation being studied, rather than to just a single trace.

\subsubsection{Transitive Closure}

Frequently, especially in the case of single bit values, many registers will 
take on the same value and Daikon
will return many equality properties. To make these properties more
manageable, Astarte takes the transitive closure of 
all the equality properties and, instead of lists of pairwise equalities,
equality properties are
presented as sets of registers that are equal. 
For example, given the three invariants 
$\mathtt{andb} \rightarrow \mathit{orig}(\mathtt{CPL}) = 3$, 
$\mathtt{andb} \rightarrow \mathtt{CPL} = \mathtt{DS\_DPL}$, and 
$\mathtt{andb} \rightarrow \mathtt{CPL} = \mathit{orig}(\mathtt{CPL})$, 
the postprocessor would return as a single property 
$\mathrm{andb} \rightarrow \{\mathit{orig}(\mathtt{CPL}), 3, \mathtt{CPL},
\mathtt{DS\_DPL}\}_=$, 
where the notation $\{\}_=$ indicates that any two signals
in the set are equal ($\forall r,s \in
\{\}_=, r = s$).

\begin{align*}
&  \mathrm{instruct} \wedge  \mathrm{precondition} \rightarrow\\
&\quad  \{\mathrm{\langle var\rangle, \langle var\rangle, \ldots}\}_=
    \end{align*}

In the next stage, properties that share a common precondition are combined to form larger properties
that more completely express processor behavior with regard to a control signal.
These properties are similar to the previous properties with the 
sole exception of having multiple sets of equal values, registers, or bits.

\begin{align*}
&  \mathrm{instruct} \wedge  \mathrm{precondition} \rightarrow\\
  &\quad  \{\mathrm{\langle var\rangle, \langle var\rangle, \ldots}\}_=\\
  &\quad  \{\mathrm{\langle var\rangle, \langle var\rangle, \ldots}\}_=\\
  &\quad  \{\mathrm{\langle var\rangle, \langle var\rangle, \ldots}\}_=\\
  &\quad \ldots
    \end{align*}

\subsubsection{OS-Specific Values}

In some cases, general purpose registers take on a particular value or set
of values for an
operating system. These values may differ across
operating systems, but there is an underlying pattern that is upheld across
operating systems and that is critical to security. For example, values must be
word aligned or in a canonical form. To identify these properties
the post-processor applies a bit
mask to equalities between values and general purpose registers to find which
bits are held constant in multi-bit registers.

\subsubsection{Identify Global Properties}
\label{sec:designglobals}
As a final step Astarte ensures that all properties
are specific to a control signal by comparing against global properties.
Recall that Astarte identifies control signals of interest in the first phase. Eleven of the 24 identified signals
were found to produce properties specific to those bits.
The remaining 13 signals all preconditioned the same global properties. During
postprocessing Astarte removes any of these global properties from the sets
of properties produced for each of the 11 control signals. These
properties are necessarily not specific to a control signal since they have been found to 
hold globally. 

\section {Evaluation}
  
I evaluate Astarte on its ability to find security properties of
the x86 architecture to answer
the following research questions:
\begin{enumerate}
	\item Can Astarte efficiently generate
assertions to prevent known CPU security bugs, specifically Memory Sinkhole
and SYSRET privilege escalation? 
	\item How effective is control signal partitioning in achieving 
    effective security properties, specifically versus manually
    discovered properties?
    \item Does Astarte produce a manageable number of properties?
    \item How can we expect the Astarte iterative mining technique
    to improve performance compared to a hypothetical approach 
    of using exhaustive precondition
    enumeration in Daikon over the same trace set?   
\end{enumerate}

The experiments are performed on a machine with an
Intel Core i5-6600k (3.5GHz) processor with 8 GB of RAM.

\subsection{Trace Data}
To avoid capturing only properties enforced by, or relevant to, a
specific operating system I generate trace data while booting multiple
operating systems. I boot two Linux
distributions (Ubuntu and Debian), Solaris, seL4, and FreeDOS ODIN. 

To achieve high instruction coverage I use Fast PokeEMU
\citep{yan2019FastPokeEMU}, a tool for testing consistency between
hardware and the QEMU emulator. 
Fast PokeEMU repeatedly executes an instruction 
with varying inputs to achieve high path coverage within an instruction with high
probability without relying on manual test generation. I execute these
instructions on the ``bare metal'' QEMU emulator.

Over all traces, 
Astarte modeled 333 distinct instructions while the Intel specification
describes 611. Reviewing the specification I find that of the 278 instructions 
not modeled
by Astarte over the trace set, 275 fall into one of five
categories: AES and SHA acceleration, mask register operations, packed value
operations, and vector operations (see Table~\ref{tab:unmodelled}). 

\begin{table}[t]
\centering
  \begin{tabular}{llr}
\toprule
  Instruction  & Description & Number of  \\ 
  Mnemonic & & Instructions \\ 
\midrule
  aes & AES acceleration & 6 \\
  k & mask register operations & 13 \\
  p & packed value operations & 87 \\
  sha & SHA acceleration & 7 \\
  v & vector operations & 162 \\
\bottomrule
\end{tabular}
\caption{QEMU unmodelled instructions by mnemonic}
\label{tab:unmodelled}
\end{table}


I analyzed 10.2GB of trace data comprising
4.1 million instruction executions. This trace
volume is consistent with requirements to find complex conditions in prior art: 
\cite{Amit2015Virtual}
found that in fewer than 1k iterations of 
tests of 4096 instructions---a similar trace volume---most known complex
race conditions could be found.

\subsection{Control Signals}
Of the 24 control signals identified prior to mining
(Section~\ref{sec:idsignals}), 11 govern a class of properties preconditioned on
that signal. The remaining 13, when used as a precondition, produced only
properties common to all preconditions; in other words, they do not govern a
particular set of behaviors. Table~\ref{tab:signals} shows the 11 control
signals along with their common name and a brief description.

\begin{table*}
\begin{tabular*}{\textwidth}{lllll}
\toprule
   Signal & Flag & Name & Description & \\
\midrule
   CS[13] & CPL & Current privilege level & Gives current ring while in protected mode \\
   SMM & SMM & System Management Mode & If set, processor is in SMM (ring -2) \\
   EFL[6] & ZF & Zero Flag & Indicates zero result of arithmetic \\
   EFL[9] & IF & Interrupt enable flag & Enables hardware interrupts \\
   EFL[11] & OF & Overflow Flag & Indicates overflow result of arithmetic \\
   CR0[0] & PE & Protected Mode Enable & If set, processor is in protected mode \\
   CR0[1] & MP & Monitor co-processor & If CR0[0]=CR0[1]=CR0[3]=1 \\
   & & & (F)WAIT raises an \#NM exception \\
   EFL[4] & AF & Adjust Flag & Indicates arithmetic carry or borrow over \\
   & & &  four least significant bits \\
   CS &  & Code Segment & The currently used program code segment \\ 
   SS &  & Stack Segment & The currently used program stack segment  \\ 
   DS &  & Data Segment & The currently used program data segment \\ 
\bottomrule
\end{tabular*}
\caption{Control signals discovered by Astarte}
\label{tab:signals}
\end{table*}

\subsection{Achieving Manageable Numbers of Properties through Postprocessing}

This section address the research question ``Does Astarte produce 
a manageable number of properties?''

At the end of the property mining phase (Sec.~\ref{sec:propertymining}), 
Daikon produces 13,722,294 properties across all instructions and preconditions.
After taking the intersection of properties across distinct trace sets and
taking the transitive closure of properties,  Astarte is left with 122,122
properties.
Identifying the global properties reduces the total to 1,393
properties, a reduction of close to five orders of magnitude from the naive
property total. These properties average 6 implied clauses each per precondition. 
Each class of
properties, defined by a single preconditioning control signal, 
has 127 properties on average.
The distribution of the number of properties and average property
size by control signal is shown in Table~\ref{tab:modesize}.

\begin{table}
  \centering
\begin{tabular}{llrrrr}
\toprule
Bit/Reg & Flag & Clauses & Properties & Clauses per Property & \\
\midrule
   CPL & CPL & 235 & 59 & 4.0 \\
   SMM & SMM & 335 & 60 &  5.6 \\
   EFL[6] & ZF & 1182 & 286 & 4.1 \\
   EFL[9] & IF & 1102 & 164 & 6.7 \\
   EFL[11] & OF & 390 & 46 & 8.5 \\
   CR0[0] & PE & 1159 & 173 & 6.7 \\
   CR0[1] & MP & 777 & 68 &  11.4 \\
   EFL[4] & AF & 1402 & 244 & 5.7 \\
   CS & CS & 465 & 55 & 8.5 \\ 
   SS & SS & 432 & 52 & 8.3 \\ 
   DS & DS & 480 & 50 & 9.6 \\ 
\midrule
   Total & & 8571 & 1393 & 6.2 \\
\midrule
   Globals & & 4187 & 246 & 17.0 \\
\bottomrule
\end{tabular}
\caption{Astarte properties' implied clauses per control signal.}
\label{tab:modesize}
\end{table}

\subsection{Historical Exploits}
\label{subsec:known_bugs}

This section addresses the research question ``Can Astarte efficiently generate
assertions to prevent known CPU security bugs, specifically Memory Sinkhole
and SYSRET privilege escalation?''

To evaluate the efficacy of the Astarte framework in producing properties relevant
for security I consider two case studies, Memory 
Sinkhole~\citep{Domas2015Sinkhole} and SYSRET privilege 
escalation~\citep{Xen2012SYSRET}.
Both of these cases received considerable attention from the security and
research communities
and, thanks to these efforts to reverse engineer the
bugs, I have information about the technical
details of the bugs beyond the high-level information provided by Intel's errata
documents.
For each case study I
examine whether the properties generated by
Astarte define secure behavior with respect to these exploits.
Table~\ref{tab:casestudyl} presents the results.
\begin{table}

  \centering
\begin{tabular}{lllll}
\toprule
No. &  Property & Found & Ctrl & Astarte\\
&&&Signal&Property \\
\midrule
   1 & CALL $\rightarrow$ SMM=0 & \checkmark & SMM & G5 \\
   2 & SYSRET $\rightarrow$ canonical(ECX) & \checkmark & CPL & 5, 7 \\
\bottomrule
\end{tabular}
\caption{Astarte performance versus known historical bugs}
\label{tab:casestudyl}
\end{table}

\subsubsection{Memory Sinkhole}
At Black Hat 2015, ~\cite{Domas2015Sinkhole}
disclosed the Memory Sinkhole escalation vulnerability in SMM.
The vulnerability allows an OS-level attacker to enter System Management Mode 
and execute arbitrary
code. The attack relies on using the $\mathtt{call}$ instruction with a
particular parameter while in SMM. 
The security properties discovered by Astarte would disallow this exploit. The
generated
properties prohibit the execution of the $\mathtt{call}$ instruction while in
SMM, capturing the secure usage of SMM in practice across the studied OSs.

\subsubsection{SYSRET Privilege Escalation}


This vulnerability, as described by the Xen Project~\citep{Xen2012SYSRET}
 arises from the way in which Intel processors 
implement error handling in their version of AMD's SYSRET instruction. 
If an operating system is written according to AMD's specification, but run on 
Intel hardware, an attacker can exploit the vulnerability to write to 
arbitrary addresses in the operating system's memory. 

The crux of the vulnerability has to do with when the Intel processor checks
that, when returning to user mode, the address being loaded into the $\mathtt{RIP}$ register from the
$\mathtt{RCX}$ register is in canonical form. Astarte generates properties that
require $\mathtt{RCX}$ to always be in canonical form when the current privilege
level is elevated, which would prevent the vulnerability. It is interesting to note that Astarte only finds this property
over traces produced by operating systems, an indication that this desired
behavior is not enforced by the hardware and must be enforced by an operating
system, as is indeed the case.

\subsection{Manually Developed Properties}

This section address the research question 
``How effective is control signal partitioning in achieving 
    effective security properties, specifically versus manually
    discovered properties?''

\subsubsection{Evaluating Astarte Coverage}

\begin{table*}
\begin{tabular*}{\textwidth}{lllp{4.4cm}}
\toprule
No. &  Property & Signal & Astarte Property \\
\midrule
   1 & IN/OUT/INS/OUTS $\rightarrow$ IOPL $\geq$ CPL & CS[13] & G65, G68-71, G104-107, G243-244 \\ 
   2 & !(JMP/CALL/RET/SYS*) $\rightarrow$ CS=orig(CS) & CS & 29 properties in [298,351] \\ 
   3 & POPF/IRET \& !CPL=0 $\rightarrow$ EFL\_13=orig(EFL\_13) & EFL[13] & G72, G111-G112, G206 \\ 
   4 & STI/CLI \& CPL $>$ EFL\_13 $\rightarrow$ EFL\_9=orig(EFL\_9) & EFL[9]  & G53, G141 \\ 
   5 & IRET \& EFL\_9=orig(EFL\_9) \&  & EFL[9]  & 1044-1045 \\ 
   & CPL $>$ EFL\_13 $\rightarrow$ EFL\_9=orig(EFL\_9) \\
   6 & IRET \& CPL$\neq$ 0 $\rightarrow$ EFL\_13=orig(EFL\_13) & CS[13] & G72, G206\\ 
   7 & SYSEXIT $\rightarrow$ CPL=0 & CS[13] & 3, 7 \\
   8 & SYS* $\rightarrow$ CPL $\leq$ DPL & CS[13]  & 37, 39 \\ 
   9 & SYS* $\rightarrow$ CS\_DPL $\leq$ CPL & CS[13]  & 37, 39 \\ 
   10 & JMP/CALL(FAR) \& CS$\neq$orig(CS)$\rightarrow$ DPL = CPL & CS & 15, 16, G18 \\ 
   11 & JMP/CALL(FAR) \& CS$\neq$orig(CS)$\rightarrow$ DPL $\leq$ CPL & CS  & 15, 16, G18  \\ 
   12 & CALL(FAR) \& CS$\neq$orig(CS)$\rightarrow$ CPL $\leq$ DPL & CS & 15, 16 \\ 
   13 & JMP(FAR) \& CS$\neq$orig(CS)$\rightarrow$ DPL $\leq$ CPL & CS  & G18 \\ 
   14 & JMP/CALL(FAR) \& CS$\neq$orig(CS)$\rightarrow$ & CS  & 15, 16, G18 \\
   &  CS\_DPL $\leq$ orig(CPL) \\ 
   15 & RET \& CS$\neq$orig(CS) $\rightarrow$ CS\_DPL$\geq$ CPL & CS & G35, G72, G90, G120, G206, G224 \\ 
   16 & SS$\neq$orig(SS) $\rightarrow$ SS\_DPL=CPL & SS & G245 \\ 
   17 & DS$\neq$orig(DS) $\rightarrow$ DS\_DPL$\geq$CPL & DS & G245 \\ 
   18 & CS\_11=1 $\rightarrow$ CS\_12=0 & CS & G245 \\ 
   19 & SS\_9=1 \& SS\_12=SS\_11=0 & SS & G245 \\ 
   20 & DS\_9=DS\_11=DS\_12=1 & DS & G245 \\ 
   21 & IRET \& EFL\_13 $\rightarrow$ CS\_DPL $\leq$ SS\_DPL & CS[13] &  G35, G72, G90, G120, G206, G224 \\
\midrule
   22- & SYSENTER/SYSEXIT \& CR0\_0=1$\rightarrow$ & CS[13] & 37, 39 \\ 
   23 & {CS=val,EIP=val,SS=val,SP=val} \\
\midrule
   24- & Properties over unimplemented  & - & unknown \\ 
   29 & MSRs or VMX instructions \\
\bottomrule
\end{tabular*}
\caption{Astarte performance versus manually specified properties.}
\label{tab:manual}
\end{table*}

\cite{Brown2017Thesis} manually studied the 
Intel 64 and IA-32 Architectures Software Developer 
Manuals~\citep{Intel2020SDM}
and crafted 29 
properties they found to be critical to security. The Astarte
properties cover 23 of the 29 manually written properties.
The remaining 6 properties required exercising processor state unimplemented in QEMU.
These properties are presented in Table ~\ref{tab:manual}.

In this case, the coverage versus manual efforts suggests Astarte
achieves coverage over security properties discovered manually.

\subsubsection{Effectiveness of Control Signals}

Manually developed properties also provide a helpful point of comparison to assess
the control signal partitioning methodology.

In Table ~\ref{tab:manual}, the relevant control signal 
on which the manually written property is conditioned is listed in column three.
Astarte discovered 11 unqiue control signals, of which 7 were used in this
portion of the evaluation. The two most commonly used control signals, the CPL
flag also denoted as CS[13] and the entire CS register, were used for 8 properties
each. Of note, the CS[13] was explored using the preconditions
described in Table~\ref{tab:expartition} to capture fixed values and value changes.
The full CS register was not compared to fixed values, only value changes.
An example property over CS value changes is shown in Figure~\ref{fig:examplepropertyB}.

Of all these properties, only those numbered 19 and 20 do not include an implication and therefore could have been discovered without specifying a Daikon precondition.

\subsubsection{Implications of Postprocessing}

In Table ~\ref{tab:manual}, 
Astarte Properties listed with a `G' prefix denote those
that are found in Daikon output under a heading that does not specify both 
an instruction and a precondition. 
These contain the ``Global Properties'' addressed
in Section~\ref{sec:designglobals}, are
identified as ``Globals'' in Table~\ref{tab:modesize}, and
are counted separately from the 1393 properties not prefixed with `G'.

For example, in the case of properties numbered 19 and 20, these properties
were discovered when mining with control signal conditions but not mining
within specific instructions.
Property number 1 offers a similar case with
respect to precondition rather than instruction,
where the relevant terms were found specified only within a particular instruction
without regard for an additional precondition over control signals.

In each case, with a different postprocessing implication, the relevant terms
in the `G' designated properties could be output in more specific properties
specifying both instruction and precondition, 
but to do so would expand the size of each property and may, as in the 
case of properties numbered 19 and 20, require a property for every instruction in the design,
greatly complicating validation efforts. These properties were not initially 
considered within output, but were added to consideration motivated by
indices 19 and 20, and I found they 
greatly eased the process of comparison to manual properties.
Astarte property G245 in particular, the property that combines all instructions to assess
invariants across the design preconditioned only on control signals but not
on instructions, captures all five properties indexed 16 to 20, and G245 describes
three control signals over 333 instructions, setting a lower bound of 333
instruction specific properties to describe the behavior captured by G245 in these
cases.

\subsection{Performance Expectations and Daikon}
\label{sec:EvalOptim}

This section address the research question 
``How can we expect the Astarte iterative mining technique
    to improve performance compared to a hypothetical approach 
    of using exhaustive precondition
    enumeration in Daikon over the same trace set''

Generating the trace data took
approximately 8 hours, processing the traces to make them suitable for Daikon took 57
minutes, identifying control bits on which to partition the property space took
44 minutes. Mining along all 
preconditions took approximately 16 hours with each
control bit costing roughly 44 minutes. Overall, the Astarte framework completed
the property generation in 29 hours running sequentially but could be accelerated
considerably by parallelizing trace generation and mining runs.

I completed the full mining process for 24 control signals as
preconditions. Excluding unused control signals from consideration provided
an estimated speedup of 5.82x.
This speedup represents a theoretical comparison to a specification mining
approach not using the Astarte process and is comparable  
to running Daikon using the minimal Astarte features necessary to encode
generated traces for property generation. However,
this implementation
 would still include such 
features as derived descriptor privilege levels and opcode aware instruction
differentiation for jump and call instructions. Running Daikon
without these features would not produce comparable properties. 
For example, each of property numbered 10 through 14
specifically applies to far jumps and calls.

This estimate assumes that control bits all take roughly the same 
amount of time to mine. In practice all mining runs specifying a 
precondition completed in around 44 minutes,
and all mining runs completed within 20 seconds of of each other.
I believe it reasonable to assume
this timing trend would apply to other untested preconditions, especially as a 
majority of these tests were performed over the 13 preconditions that were not
found to be associated with any unique design behavior, the expected case for
the unstudied preconditions.

\subsection{Operating System-Enforced Properties}
\label{sec:OS}

\begin{figure}
\centering
\includegraphics[width=\columnwidth]{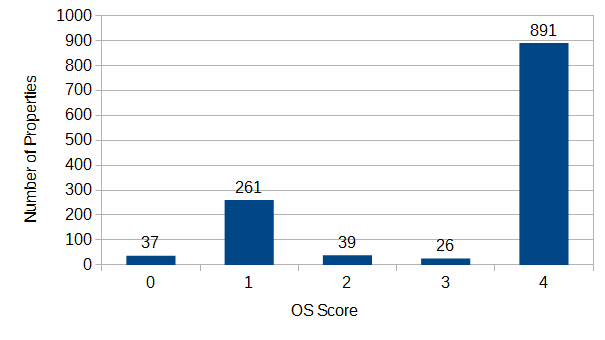}
\caption{Distribution of properties by number of enforcing OSs}
\label{fig:osscore}
\end{figure}

\begin{figure}
\centering
\includegraphics[width=\columnwidth]{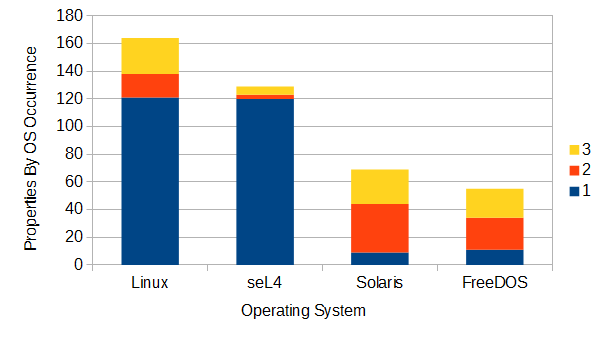}
\caption{Distribution of partially OS enforced properties by enforcing OS}
\label{fig:osoverlap}
\end{figure}

When mining over traces from different operating systems,
some properties are found over all operating
systems and some over only a subset of operating systems
or only on bare metal traces with no operating system. In
Figure~\ref{fig:osscore} I show how many operating systems are found to
enforce each property.  
Figure~\ref{fig:osoverlap} shows, for each property enforced by 1, 2,
 or 3 operating systems, which operating systems enforce these properties.
 
Properties were predominantly
 enforced either by a single operating system or by all operating
 systems. Properties enforced by a single operating system are likely 
 fall into two main possibilities:
either the properties are well-founded properties that, when enforced,
make the operating system more secure in some way, or that they are false positives
and found only within a single operating system for this reason.  In manual inspections of 
properties, I found that many of the properties unique to Linux and
seL4 were related to ensuring the safety of the 
specific implementations of system calls used by the operating system.
Unsurprisingly, more properties were enforced on Linux and seL4 which
have the highest usage levels and most rigorous theoretical assurances
respectively. The remainder of unique properties governed specific instruction
usage from specific processor states only exercised by that operating system that
may or may not be associated with security.  

I interpret properties
enforced by all operating systems to be necessary implementation features
as changing any one of them would would likely cause compatibility issues
across many operating systems. This assessment is extensible
to properties implemented by all but one operating system,
especially as the operating system most frequently missing was seL4.
As seL4 by design has provably correct behavior it cannot
rely on undocumented or incidental features.  Without the burden of provable
correctness and security enforcement, other operating systems may make reasonable
assumptions of processor behavior. These assumptions may eventually become part
of the processor specification if many operating systems come to rely on
them, making it difficult for hardware designers to modify the expected, though
undocumented, behavior.

The few properties enforced by just two operating systems usually govern
behavior of a very specific type of system call that is enforced
by precisely two operating systems. A few properties
 govern specific instruction usage enforced by precisely
two operating systems.  Similarly, these may be best practice
or false positive properties, but represent a small minority of output.

There were also a few properties found to be enforced
on bare metal traces but not operating system traces.
I regard these as either false positives 
or these are vestigial properties that persist in hardware but
OSs no longer need to rely upon.

\subsection{Properties in the Specification}

\begin{figure}
\centering
\includegraphics[width=\columnwidth]{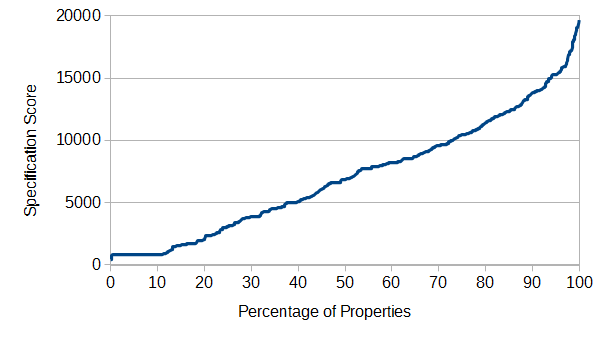}
\caption{Properties by relevant mentions in Intel documentation}
\label{fig:propscore}
\end{figure}

To provide a sense for how difficult 
the properties generated by Astarte would be to find manually,
I use a scoring function for properties that
considered each bit or register within a property against
how many times that bit or register is referenced in the Intel Software
Developers Manuals~\citep{Intel2020SDM} to give a sense of how many pieces of discrete
information must be considered to generated a property. Figure~\ref{fig:propscore} shows
the cumulative distribution function of this specification score for properties. 

\begin{figure}
\centering
\includegraphics[width=\columnwidth]{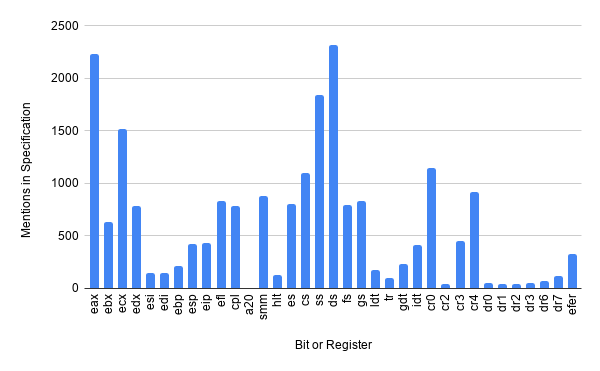}
\caption{Signals by mentions in Intel documentation}
\label{fig:specscore}
\end{figure}

Properties typcially would require reviewing approximately 7000 mentions (median 6874, mean 7088)
 with a minimum of 413, a maximum of
19669, and about 8.9 million in total.  The distribution is nearly uniform
with slight clustering at the minimum and slightly longer tails on the
maximum. Figure~\ref{fig:specscore} shows how many discrete mentions of each bit or
register occur in the ISA specification. 

\section{Limitations}

In this section, I will discuss the threats to validity for
properties produced using Astarte, including false positives
and false negatives.

With regard to false positives, 
Daikon's inference engine infers only properties 
not violated over the trace set,
so using Daikon's inference engine within Astarte 
restricts false positives to two cases: 
false positives arising from trace generation, which can include
limitations of emulation, and false positives arising from
misclassification, which can occur when functional properties
are classified as security properties. Some examples of false positives
are shown
in Section~\ref{sec:functionalprops}, along with 
a discussion of consequences
and potential remedies.
Though presented within the evaluation,
Section ~\ref{sec:OS} also discusses possible 
false positives in the context
of properties solely over bare metal traces and over
no operating systems, which gives a false positive rate
of 37 out of 1393 or approximately 2.7\%.

With regard to false negatives, 
they fall into two cases: known and unknown.
There are six known false negatives from the evaluation
which result from limitations in the emulator
and debug logging. Two false negatives relate to logging while
switching into long mode,
and four false negatives relate to logging while using VMX.
Unknown false negatives could arise from limitations in trace
coverage, in logical specificity, or in the abstraction level.
Even within ISA level conditions, Astarte may miss
any property not expressible in the grammar which specifies
preconditions over single control bits, or
 properties involving individual bit flags of a register that was not decomposed into derived variables.

\subsection{Trace Reliance}

As with any specification mining technique, Astarte many only determine
invariants that hold over traces. In the case of traces over buggy hardware,
discovered invariants may form a specification describing buggy behavior.
Traces may not cover all cases that can be reached by a design or even
occur during normal design operation.

For example, within the trace set, Astarte only observed usage of System Management
Mode while running traces using the seL4 operating system.
SeL4 offers much higher degrees of security assurances than other operating systems,
so the corresponding properties in SMM may restrict what could be allowable behavior
to ease the burden of making security assurances.

None of AES, SHA, mask register
operations, packed value operations, or vector operations were used at
any point in the trace set. 
Consequently, Astarte does not define secure behavior with respect to any of these instructions, and this lack of specificity is a threat to the validity of the output specification.

\subsection{Emulation Reliance}

Astarte relies on traces produced in emulation and logged through debug ports.
Further, traces are produced in translation blocks rather than on a per-instruction
level, while properties target instructions. While QEMU does support
single instruction translation blocks, doing so incurs memory and I/O costs so
prohibitive as to preclude trace generation
covering interesting design behavior, such as completing a Linux boot, which
crashes early due to memory requirements when attempting to log each instruction.

Further, emulation may not precisely capture hardware implementation.
Astarte assumes the correctness of emulation. 
Prior work, such as Fast PokeEMU ~\citep{yan2019FastPokeEMU},
addresses directly the correctness of CPU emulation.

QEMU is capable of emulating VMX instructions
 but due to implementation details regarding 
nested virtualization is not
able to log during VMX emulation. 
Consequently, there are no VMX instructions in the
trace set, a limitation relevant to false negatives.
Within the set of manually discovered properties, four describe
VMX.

Beyond VMX, the QEMU implementation in use in this chapter contains a known error 
regarding switching into
long mode while debugging, which was used for trace generation.
Further, long mode is implemented in the model specific EFER control
register, and similarly to VMX, processor extensions and model specific
registers are not fully supported by QEMU.
Ultimately I was unable to 
log traces while in long mode. 
While potentially related to trace limitations or human error,
this coverage gap occurs around known limitations for
the emulator.

\subsection{Functional Properties}
\label{sec:functionalprops}

Astarte
discovers a number of properties that could reasonably be considered more
closely related to functional correctness than secure behavior.

Of the 11 control signals discovered by Astarte, three strictly implement
arithmetic operations:
\begin{itemize}
\item EFL[04]: AF: the Adjust Flag
\item EFL[06]: ZF, the Zero Flag
\item EFL[11]: OF, the Overflow Flag
\end{itemize}

While each of these do affect conditional operation of the processor,
they do so largely on the basis of jump/call conditions. Ultimately,
each describes arithmetic operations upon which the processor relies
to implement secure control flow, but this is not their main purpose
within the design and they
are not necessarily security specific.

Together, properties over these signals represent 575 out of 1393 
discovered properties, a false positive rate under this
characterization of just over 41\%.
The AF and ZF together represent the
two signals with the most properties. AF in particular has 286 properties,
over a hundred more than any other signal except ZF at 244 and more
than double the average.

False positives of this nature could greatly increase 
effort required for validation.
One path to validation with respect to the Astarte specification is to use
formal tools, such as model checkers, which incur high time and memory costs
 over
large designs such as x86, and this false positive rate would almost double
these costs.

One possible rememedy to reduce the false positive rate with respect
to misclassification is to use machine learning to characterize output
properties based on known vulnerabilities, as shown by SCIFinder~\citep{Zhang2017SCIFinder}.
SCIFinder used developers’
repositories, bugtracker databases, and email forums to create
a training data set, and none of these while studying a
closed source design. However, SCIFinder demonstrates a
promising technique to for characterizing security properties
of CPUs, and could be used to
reduce Astarte's false positivity rate given adequate training data, which
is likely available to hardware designers.

\subsection{Specification Logic and Abstraction Level}

Astarte does not discover any temporal properties or information flow
properties. Properties are described at ISA level and do not address
any microarchitectural state.

Pipeline properties arise from both the 
lack of temporal properties and the ISA abstraction, and can describe
important security requirements. Hardware validation efforts such as
~\cite{Hicks2015SPECS} and ~\cite{Zhang2017SCIFinder} define
security properties at the register transfer level across pipeline
stages.

\section{Conclusion}

This chapter presents Astarte, a framework for mining security critical
properties from a closed source, CISC architecture. Astarte produces
manageable numbers of properties that capture the behavior of the processor
under security-relevant control signals. Astarte addresses issues of complexity
and closed-source designs not seen in RISC-based security specification
miners. I show that Astarte can automatically generate specifications
containing manually discovered security properties and 
capable of securing designs against historical bugs.

\chapter{Undine: Mining for Temporal Properties}
\label{chap:Undine}

\section{Introduction}

Efforts at automatic security property generation can
struggle to move beyond invariants to more powerful logics of specification,
such as those able to capture temporal properties.
Early specification miners for software and hardware, such as 
DIDUCE~\citep{Hangal2002Diduce} and
IODINE~\citep{Hangal2005Iodine} studied only invariants or subsets of invariants,
and early efforts at temporal specifications frequently addressed specific
patterns rather than temporal logics generally, like Javert~\citep{Gabel2008Javert},
or study hardware at higher abstraction
levels~\citep{Danese2015Automatic,Danese2016Automatic}.
Prior security efforts at the register transfer level (RTL) also studied invariants, such
as SCIFinder (for `Security Critical Invariant Finder'),
used statistical learning to identify invariants as security critical or
not~\citep{Zhang2017SCIFinder}. SCIFinder produced only
non-temporal properties and relied on human expertise to produce the initial
training set of properties.
To achieve automatic RTL temporal security specification, an automated approach to 
capture the security critical patterns within designs 
and a corresponding library
of such patterns must be developed. 
With such an approach, a
new design can be mined to generate a set of security properties with little 
human intervention.

To address this need with specification mining, 
I developed Undine, a tool for mining temporal security 
specifications of processor
designs. The specifications take the form of linear temporal logic formulas and
capture properties that are critical to the security of the processor.

Undine introduces the notion of event labels and uses these labels to find the 
patterns that
are common to the known security properties discovered manually
or composed from invariants over the design. For example,
many invariants may hold when restricting the considered system to only be studied
after it correctly comes out of reset, and invariants that only hold
under these certain design conditions that may be defined temporally. 
Using event labels and Undine, I build a library of labelled property
templates and develop a specification miner for use with the labelled templates. 

Undine uses a modified version of the Texada LTL Specifications 
Miner~\citep{Lemieux2015Texada}. 
Texada mines properties expressible in linear temporal logic (LTL), a modal logic
which can encode formalae about the future of paths.
Texada offers
functionality that dynamic invariant detection, such as by Daikon~\citep{Ernst2007Daikon},
does not, but requires LTL templates to be provided to the miner.
I modify Texada to accept traces containing
event labels and to reason effectively about register slices.
I added a preprocessing step to provide the needed labelling information
to traces of execution, and to apply filters to reduce the complexity of the search.
Given these output properties based on security templates,
Undine then uses a  postprocessing step to logically compose related properties 
into a concise set of more expressive properties that are critical to security.

I demonstrate the use of the library of labelled LTL security property 
templates and the Undine label-aware
specification miner by mining security
specifications of three open source RISC processors: OR1200, Mor1kx, and
RISC-V. Using the library of templates, Undine automatically mines
25 of the 28 known security critical properties on OR1200.
Undine additionally finds new security critical properties
that require temporal logic to express.
 I provide an
example exploit for one such property over the Mor1kx CPU design.

\section{Properties}

Undine generates properties related to temporal relations between elements
of design state. The precise structure of these properties is described
in detail alongside the Undine methodology in Section ~\ref{sec:undinemethod}.
In this section, I will briefly describe the reasoning behind understanding
these properties as related to security.

\subsection{Security Properties for Undine}

Undine is intended to discover properties that describe the desired
secure behavior of open source RISC designs in linear temporal logic (LTL)
over gigabyte trace sets at register transfer level (RTL).
Undine generates temporal properties as expressions over 
up to three propositional variables 
that either specify that the value of a register, a subfield, or a bit is equal to
some
constant, 
specify the delta value of
a register, or specify an equality between the values of two registers.
Within the context of the Undine framework,
the working definition of a security temporal property for a design 
is one of these
expressions for which a violation would indicate the presence
of a vulnerability that an
attacker may exploit to breach a security requirement.
In brief, for each security property, 
there is some corresponding attack.

\subsection{Evaluating Undine Properties}

There may be many security properties of a design, and since not
all attacks are known, there will be some security properties
for which the corresponding attack is not yet known.
Assessing these properties
poses a challenge,
as for some output property that does not describe
behavior preventing a known vulnerability,
it could be the case that the property is relevant to some unknown
vulnerability or that it is simply a functional property.

Ideally, there would exist complete documentation regarding
the security requirements of open source
RISC CPU designs.
In such a case, Undine's properties
could be evaluated for coverage of known requirements, 
and the relevance of the overall generated specification to
security requirements could be also be evaluated.

In practice, there is no large body of known security temporal requirements
for RISC designs. If there were, there would be little need for a tool
such as Undine to generate security properties.
Nevertheless, it is possible to study properties generated
by Undine and determine how
a design not upholding a particular property could be vulnerable to some exploit.
However, Undine produces many properties, and as a practical matter 
I could not produce corresponding attacks for all of the generated properties. 
Given this challenge, I assessed Undine in two main ways. First, I compared
Undine against known, manually specified properties from prior work~\citep{Zhang2017SCIFinder}
that could be expressed as Undine LTL expressions.
Second, I created and demonstrated an example
exploit violating a temporal property discovered by Undine. 
The exploit, on the Mor1kx design, 
manipulates status registers prior to reset in such 
a manner that
a user may gain supervisor privileges after an exception
is raised. 
This exploit shows that the violated temporal property enforces
a security agreement related to privilege levels, and
establishes this property as a security temporal
property. The exploit is discussed in greater detail in Section~\ref{sec:exploit}.

\section{Methodology}
\label{sec:undinemethod}

\begin{figure}
  \centering
  \includegraphics[width=\columnwidth]{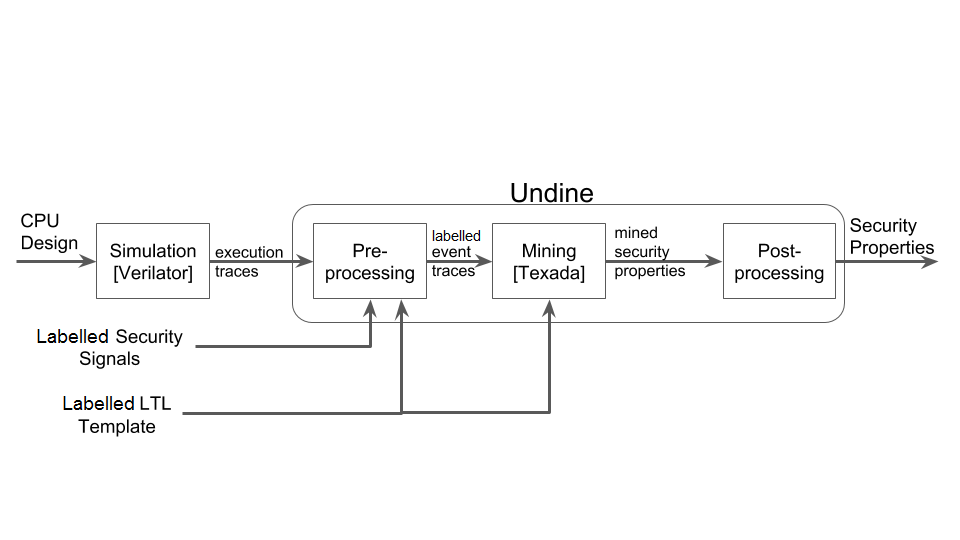}
  \caption{An overview of the Undine workflow.}
  \label{fig:undineoverview}
\end{figure}

Figure~\ref{fig:undineoverview} provides an overview of the Undine workflow. As a
preliminary step, the processor design is simulated to generate traces of
execution.  The traces are then input to Undine, which works in three steps: 
preprocessing, mining, and
postprocessing. During preprocessing Undine converts the traces of execution to
traces of labelled events and then applies a filter. During mining Undine takes
filtered, labelled event traces and a labelled property template, and produces a set of security critical
properties. During postprocessing Undine synthesizes properties to produce a
manageable set of properties that can be understood by the user and are critical
to the security of the processor.

I will use
the following security property as a motivating example while describing this process:

\begin{align*}
&  \mathrm{assert} ~  \mathrm{property} \\
  &\quad  (\lnot((\mathrm{ex\_insn } ~\&~ \mathrm{0xFFFF0000}) >> 16 == 8192))\\
  &\quad  ||~(\mathrm{id\_flushpipe } == 1);\\
    \end{align*}

In this property, the ampersand character $\&$ denotes the bitwise ``AND'' operation.
This property was developed manually by ~\cite{Hicks2015SPECS}.
It states that when a system call instruction is being
executed that the instruction pipeline should be flushed at the instruction decode phase. It is
critical to security because a system call causes a change in
privilege level, and the instruction following a syscall in the pipeline, which will not be part of the system call, should not be
executed at the elevated privilege level.

\subsection{Trace Generation}
For Undine, a \emph{trace of execution} is produced by
simulating the register transfer level (RTL) specification of the design under
consideration. The value of each signal in the RTL model is logged at
every clock cycle. More formally, a trace $T$ is an
ordered sequence of time-stamped, signal--value pairs: 
\begin{align*}
  T =&[(q,x)_t, (r,y)_t, (s,z)_t, \ldots, \\
    &(q,x')_{t+1}, (r,y')_{t+1}, (s,z')_{t+1}, \ldots,\\
    &(q,x'')_{t+2}, (r,y'')_{t+2}, (s,z'')_{t+2}, \ldots],
\end{align*}
where $q,r,s$ represent state-holding signals in the design, and $x,y,z$
represent numeric values. Tick-marks indicate the passage of a single clock cycle: if $x$
represents the value of register $q$ at time $t$, $x'$ represents the value of
register $q$ at time $t+1$. I will use this notation throughout the chapter. 
Where context makes the
meaning clear, I will sometimes overload
terms and use $q, r, s$ to mean both the register and its
value. 

Each signal--value pair in the trace is an \emph{event}. A standard value change dump (VCD) file as produced by many
simulators suffices as a trace of execution. An excerpt from the VCD file produced by simulating
the OR1200 processor is shown in Figure~\ref{lst:vcd}.

\begin{figure}
\begin{lstlisting}[frame=single]
or1200_ctrl.ex_insn == 1234
or1200_ctrl.id_flushpipe == 1
... // this denotes a change in time
or1200_ctrl.ex_insn == 4321 // etc.
\end{lstlisting}
\caption{Commented excerpt of a trace of execution from the OR1200 processor}
\label{lst:vcd}
\end{figure}

\subsection{Event Labels}
\label{sec:labelledevents}
Central to the design of Undine is the notion of a \emph{labelled
  event}. I define five event labels, which apply to events that 
  may occur at any time point within some trace.

\begin{itemize}[noitemsep,leftmargin=*]
  \item register--register ($\mathtt{RR}$): $(q_1 == q_2)$. Two registers, $q_1$
    and $q_2$, have the same value.
  \item delta--register ($\mathtt{DR}$): $(q' == q+y$). Register $q$ changes by
    some value $y$ in the next clock cycle.
  \item register--value ($\mathtt{RV}$): $(q == y)$. Register $q$ has value $y$.
  \item slice--value ($\mathtt{SV}$): $(q[i:j] == y)$. A slice of register $q$ has
    value $y$.
  \item bit--value ($\mathtt{BV}$): $q[i:i+1] == y$. The $i^{\mathrm{th}}$ bit of
    register $q$ has value $y$.  
\end{itemize}

Returning to the example, a slice of the register 
$\mathtt{ex\_insn}$ is
compared to a value, in this case using a bit mask. (This comparison checks
whether the instruction in the execute phase of the pipeline is a syscall.)
This is expressible as an equivalent
slice--value event.  The $\mathtt{id\_flushpipe}$ clause is an example of a
register--value event. The original property can be restated using labelled events.

\begin{align*}
&  \mathrm{assert} ~  \mathrm{property} \\
  &\quad  (\lnot((\mathrm{ex\_insn } ~\&~ \mathrm{0xFFFF0000}) >> 16 == 8192)) & \mathrm{//~SV~event} \\
  &\quad  ||~(\mathrm{id\_flushpipe } == 1); & \mathrm{//~RV~event} \\
    \end{align*}

Event labels inform specification mining in two ways:
\begin{enumerate}
\item Registers in the design are associated with a particular
label and will only appear in events of the correct label;
\item Property templates
are written in terms of labelled events and only property instances with the correct labelling
will match a given template.
\end{enumerate}

\subsection{Grammar of LTL Properties}

Undine mines for properties by looking for possible instantiations of a given
template. It is not limited to a predefined set of property templates, but
rather takes the template as an input from the user. A user is free to create
their own template or choose a template from the library I developed. 
Figure~\ref{fig:grammar}
defines the language of properties expressible in Undine. 
The temporal operators ($\mathbf{G,
    U, X}$) have the standard definitions of Globally, Until, neXt. The event
  labels ($\mathtt{RR, DR, RV, SV, BV}$) are as
  defined in section~\ref{sec:labelledevents}.

\begin{figure}
  \centering
\begin{align*}
  \mathit{LTL} \doteq&~ \mathbf{G}(\phi) \\
  \phi \doteq&~ \phi \rightarrow \phi ~|~ \neg\phi ~|~ \phi  ~|~ \phi \wedge
  \phi ~|~ \phi \vee \phi\\
  &   ~|~ \phi~\mathbf{U}~\phi ~|~ \mathbf{X}~\phi ~|~ e\\
  e \doteq&~ \mathtt{RR} ~|~ \mathtt{DR} ~|~ \mathtt{RV} ~|~ \mathtt{SV} ~|~ \mathtt{BV}
\end{align*}
\caption{The grammar of labelled LTL properties}
\label{fig:grammar}
\end{figure}

\subsection{Preprocessing}
\label{sec:preprocessing}
The preprocessor takes as input a set of execution traces and produces a filtered set of
labelled event traces ready for specification mining. There are two tunable
parameters to the preprocessor that determine the trace transformation:
\begin{itemize}[noitemsep]
\item labelling information for signals in the design, and
\item register slice size
\end{itemize}

In prior work, ~\cite{Zhang2017SCIFinder} found that there is a subset of registers in the
design that are associated with properties critical to
security. I dub these the \emph{security-critical
registers}. Undine extends this idea further. On manual inspection of the properties,
I noted that a
security-critical register will occur within the assertion grammar for security properties
in a predictable and consistent manner. For
example, the $\mathtt{id\_flushpipe}$ register from the running example 
would be labelled as $\mathtt{RV}$
event. In the security-critical properties in 
~\cite{Zhang2017SCIFinder} and ~\cite{Hicks2015SPECS} the
$\mathtt{id\_flushpipe}$ register appears only in events of that would be
labelled as $\mathtt{RV}$. 
The first parameter to the preprocessor is the labelling information for
the security-critical registers in the design.

Hardware designs often use bit packing, for example, storing 32 individual
control bits in a single 32-bit register. Another design tactic is to encode
semantic information in a slice of a register, as when the highest-order 16
bits of the 32-bit $\mathtt{ex\_insn}$ register determine whether 
the instruction is a systen call instruction.
Security
properties are often concerned with the control and semantic information
available at the sub-register level. To enable this, the second parameter to the
preprocessor is the register slice size: the preprocessor will break every
register into its component slices of the given size. As I discuss in
Section~\ref{sec:complexity}, register slicing situationally reduces the time 
cost of
property mining for any given template.

The trace of execution is converted to a filtered trace of labelled events
as follows. At each clock cycle in the trace, each register is split
according to the register slice size parameter. Each
register is then labelled according to the signal labelling
parameter. Next, a set of \emph{derived events} are added to each clock cycle in
the trace. The derived events are calculated as follows. For every event label that appears in the property template, for every
register of the appropriate label, the set of possible derived events is added to
the trace. The execution trace from Figure~\ref{lst:vcd} could be sliced and labelled
as follows, given certain input parameters.

\begin{figure}
\begin{lstlisting}[frame=single]
ex_insn[15:0] == 1234 // SV
ex_insn[31:16] == 0   // SV
id_flushpipe == 1     // RV
\end{lstlisting}
\caption{Sliced and labelled excerpt of a trace of execution from the OR1200 processor}
\label{lst:vcdlabelled}
\end{figure}

Finally, at each clock cycle,
registers in the execution trace that are of a label that does not correspond to
any of the event labels in the property template under consideration are removed
from the trace.

\subsection{Property Mining}

After preprocessing, the filtered labelled event traces and the property template
are passed to the specification miner.

Undine builds on the Texada LTL Specifications Miner~\citep{Lemieux2015Texada}.
Texada takes in a trace of events and a property
template and produces all property instantiations of the given 
template that are true of the event trace.
I modify Texada to handle labelled events and labelled LTL property templates. With event labels, potential properties that would otherwise match the
property template but have a label mismatch can be discarded early. The event
labels provide an effective filter at both the preprocessing and the mining
stage.

Two additional modifications to Texada include discarding registers with
uninitialized values and adding support to recognize and effectively handle
sliced registers.

 \subsection{Postprocessing}

The postprocessing step combines and simplifies properties to produce a more
manageable set of final properties. First, all properties that contain an
implication are sorted by antecedent. Properties that have the same
antecedent are grouped into a new property in which all the consequents are joined
together in a conjunction. Second, properties containing an implication are sorted by consequent
and all properties with the same consequent are grouped into
a new property in which all the antecedents are joined together
in a disjunction. Finally, the
properties are simplified using the Z3 SMT solver~\citep{DeMoura2008Z3}.

 \subsection{Complexity}
 \label{sec:complexity}
As with most specification miners the time complexity of Undine is exponential in
the number of unique terms in the property template under 
consideration~\citep{Gabel2008Symbolic}. Undine's complexity is given
by $e^T$, where $e$ is the number of unique events in the set of
traces being mined and $T$ is the number of events in the
template.

Register slice size
affects the run time of Undine in two ways. If the slice size is aligned with
semantic and control components of a register the templates required to capture
the desired security properties tend to be simpler. If, on the other hand, the
slice size is too big or too small the number of unique events in the template ($T$)
grows. On the other hand, smaller slices always reduce the number of unique
events in a trace ($e$).

\section{Evaluation}

I evaluate Undine and the library of labelled templates
 on its ability to extract temporal security properties at
the register transfer level to answer
the following research questions:
\begin{enumerate}
    \item 
Can Undine find known properties formulated as LTL expressions?
    \item 
Can Undine find new temporal properties that secure designs against 
potential exploits?
\end{enumerate}

In comparison to other tools, a helpful point of comparison would be to use
Undine
trace generation and encoding stages but interface with Texada directly 
rather
than using the Texada variant modified to apply the Undine labelling system
during property generation.
In this sense,
Section ~\ref{sec:vsTexada} offers a point of comparison for the 
Undine mining stages compared to Texada.

\subsection{Property Templates}
I developed a library of nine labelled LTL templates that describe the patterns
common to security critical properties for open source, RISC, pipelined
processors. These are described in Table~\ref{tab:u1}. The first eight
templates in the library come from studying security critical properties
developed, either manually or semi-automatically, by ~\cite{Zhang2017SCIFinder}
and ~\cite{Hicks2015SPECS}. 
The ninth template comes from my own
study of the processor design specifications.

The third and fourth columns of Table~\ref{tab:u1} list how many properties
each template produced when mining the OR1200 processor
(Section~\ref{sec:traceslength} provides details on the evaluated designs and
mining configuration). 
Column five lists how many of the known security critical properties of
prior art were found
by each template. In total, the library of templates covers 25 of the 28
security critical properties of prior
work. The 3 
properties not found require a bit shift that is determined dynamically, which
is not supported by the Undine grammar.

Template 9 uniquely uses the `$\mathbf{U}$' (until) LTL operator and is necessary for finding
properties that ensure the processor is initialized correctly. I discuss this
template further in the next Section~\ref{sec:temporaltemplates}.

\begin{table}[t]
  \centering
  \begin{tabular}{llrrr}
   & Labelled & Mined & Postprocessed & Known\\
ID & Template & Properties & Properties & Properties\\
\midrule
1 & $\mathbf{G}(\mathtt{RR}_a$) & 2 & 2 &1 \\
\midrule
2 & $\mathbf{G}(\mathtt{SV}_a \rightarrow \neg\mathtt{RR}_b$) &  46843 & 32 &2 \\
\midrule
3 & $\mathbf{G}(\mathtt{SV}_a \rightarrow \mathtt{SV}_b$) & 8134 & 376 &2 \\
\midrule
4 & $\mathbf{G}(\mathtt{SV}_a ~|~ \neg\mathtt{SV}_b$) & 5794 & 431 &1 \\
\midrule
5 & $\mathbf{G}((\mathtt{SV}_a ~\land~ \mathtt{SV}_b) \rightarrow \mathtt{RR}_c$) &  1026262 & 19 & 14 \\
\midrule
6 & $\mathbf{G}(\mathtt{SV}_a \rightarrow \mathtt{DR}_b$) &  13088 & 4 &1 \\
\midrule
7 & $\mathbf{G}((\mathtt{SV}_a ~\land~ \mathtt{SV}_b) \rightarrow \mathtt{BV}_c$) &  204138 & 3 &1 \\
\midrule
8 & $\mathbf{G}(\mathtt{SV}_a \rightarrow (\mathtt{SV}_b ~|~ \mathtt{RR}_c$)) & 525322 & 648 &1 \\
\midrule
9 & $\mathtt{RR} ~\mathbf{U}~ \mathbf{G}(\mathtt{BV}$) & 134 & 134 & 0 \\
 \midrule
  \end{tabular}
  \caption{LTL templates over labelled events}
\label{tab:u1}
\end{table}

\subsection{Mining with Temporal Templates}
\label{sec:temporaltemplates}

In prior work, properties are often defined for the processor starting
at the first clock cycle after reset. These assume that processor state is
initialized correctly; if it is not, security may be compromised without
violating any property. Specifying the sequence of events required for
secure initialization requires temporal operators or some equivalent formulation, 
a prospect not addressed in prior work on
defining security properties.

Using the ninth template `$\mathtt{RR} ~\mathbf{U}~ \mathbf{G}(\mathtt{BV})$', 
Undine mines properties on the Mor1kx processor and finds seven
groups of registers that must be set as equal to each other until the initialization
period has ended (until $\neg\mathtt{reset}$). These properties are listed in
Table ~\ref{tab:untilprops}. The first six properties describe registers that are free to
change their values after reset; the last property describes registers that
must always be equal and could therefore have been captured with the simpler
$\mathbf{G}(\mathtt{BV})$ template. In Section~\ref{sec:exploit} I use
the first property listed in Table~\ref{tab:untilprops} as a
case study and examine how a design that violates of the property 
may contain an exploitable security vulnerability.

\begin{table*}
\begin{tabular*}{\textwidth}{llll}
Registers Equal During Initialization & Description \\
\midrule
  spr\_esr, spr\_sr, spr\_sr\_o  & Status registers and exception status register  \\
\midrule
  ctrl\_epcr\_o, pc\_execute\_to\_ctrl & Program counter and exception program counter  \\
\midrule
 ctrl\_lsu\_adr\_o, dbus\_dat\_o, du\_dat\_o, & \\
  mfspr\_dat\_o, pc\_decode\_to\_execute,   & Decoder and data channels  \\
 pc\_fetch\_to\_decode, spr\_bus\_dat\_i & \\
 spr\_bus\_dat\_o\\
\midrule
  decode\_rfa\_adr\_o, decode\_rfb\_adr\_o  & Decode stage register file address registers \\
  decode\_rfd\_adr\_o \\
\midrule
  fetch\_rfa\_adr\_o, fetch\_rfb\_adr\_o & Fetch stage register file address registers  \\
  wb\_rfd\_adr\_o\\
\midrule
  ctrl\_rfd\_adr\_o, execute\_rfd\_adr\_o  & All pipeline register file data registers  \\
 fetch\_rfd\_adr\_o, wb\_rfd\_adr\_o \\
\midrule
  du\_dat\_i, snoop\_adr\_i  & Debug ports to databus (globally true) \\
\midrule

\end{tabular*}
\caption{Properties mined using initialization template on Mor1kx.}
\label{tab:untilprops}
\end{table*}

\subsection{Labelling and Performance vs. Texada}
\label{sec:vsTexada}

I examine the performance benefits of introducing labelled events,
including a direct comparison with baseline Texada to contextualize
Undine within prior art.
Table~\ref{tab:impls} compares mining time for each template for each of
three versions of Undine: Texada, Security Signal, and Labelled. 
The Texada implementation uses traces containing events from all registers,
allows any signal to be associated with an event of any 
label, 
and does not include label checking in the miner. 
This is termed Texada because it is equivalent to
using the Undine workflow only for trace generation and encoding
and thereafter using an unmodified Texada implementation.
In this sense, the Texada implementation represents a helpful 
comparison to prior work on template driven LTL specification
mining, with the caveat that other Undine techniques, such as register
slicing~\ref{sec:slicing}, still inform performance
and feasibility.
The Texada implementation reaches a four hour timeout in all cases,
including the minimal template $\mathbf{G}(\mathtt{RR}_a$), 
which as a point of comparison could be mined using invariant
detection, such as through Daikon~\citep{Ernst2007Daikon},
rather than an LTL miner.
Security Signal implementation uses traces with only 
registers associated with security critical 
properties, allows any signal to be associated with an event of any 
label, and does not include label checking
in the miner. The Labelled implementation uses traces with only 
registers associated with security critical 
properties, uses only events in which signals have the correct label to
be security relevant, 
and includes
label checking in the miner. In all except the most trivial cases, 
mining is prohibitively expensive
without any labelling, and in the cases of the fifth, seventh, and eighth template,
mining also timed out at four hours in the security signals case.

\begin{table}[t]
  \centering
  \begin{tabular}{llrrr}
   & Labelled & Texada & Security & Labelled\\
ID & Template & Miner & Signals & Templates\\
\midrule
1 & $\mathbf{G}(\mathtt{RR}_a$) &t/o & 6.823 & 0.600 \\
\midrule
2 & $\mathbf{G}(\mathtt{SV}_a \rightarrow \neg\mathtt{RR}_b$) &t/o & 38.682 & 1.777 \\
\midrule
3 & $\mathbf{G}(\mathtt{SV}_a \rightarrow \mathtt{SV}_b$) &t/o & 122.186 & 7.826 \\
\midrule
4 & $\mathbf{G}(\mathtt{SV}_a ~|~ \neg\mathtt{SV}_b$) &t/o & 71.694 & 6.856 \\
\midrule
5 & $\mathbf{G}((\mathtt{SV}_a ~\&~ \mathtt{SV}_b) \rightarrow \mathtt{RR}_c$) &t/o &  t/o & 217.988 \\
\midrule
6 & $\mathbf{G}(\mathtt{SV}_a \rightarrow \mathtt{DR}_b$)  &t/o & 122.186 & 18.445 \\
\midrule
7 & $\mathbf{G}((\mathtt{SV}_a ~\&~ \mathtt{SV}_b) \rightarrow \mathtt{BV}_c$) &t/o &  t/o & 515.168 \\
\midrule
8 & $\mathbf{G}(\mathtt{SV}_a \rightarrow (\mathtt{SV}_b ~|~ \mathtt{RR}_c$)) &t/o &  t/o & 1521.896 \\
\midrule
9 & $\mathtt{RR} ~\mathbf{U}~ \mathbf{G}(\mathtt{BV}$) &t/o & 27.995 & 0.987 \\
\bottomrule
\\
  \end{tabular}
\caption{Time in seconds to mine the template library by miner implementation}
\label{tab:impls}
\end{table}

\subsection{Slicing and Performance}
\label{sec:slicing}

\begin{figure}
  \centering
  \includegraphics[width=\columnwidth]{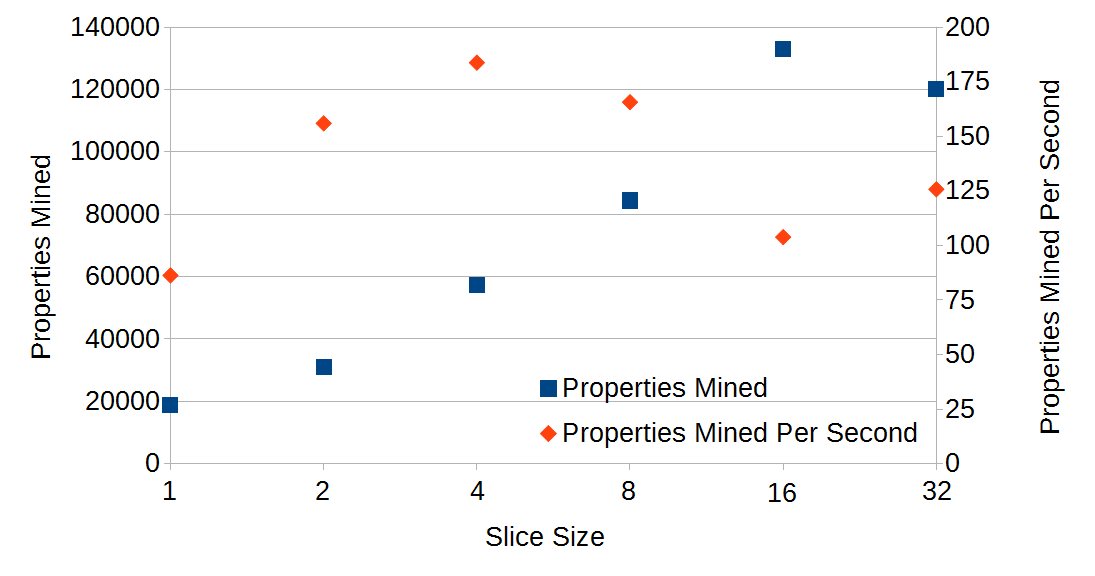}
  \caption{Mining rate and quantity of output properties by slice size}
  \label{fig:2axis}
\end{figure}

The register slice size is a parameter to the preprocessor and is adjustable by
the user. Smaller slice sizes lead to fewer possible unique events for a given
trace, giving a performance boost to the miner. However, changing the slice size
in either direction can affect the number of property instantiations for any
given template as well as the rate at which properties can be mined. 
Figure~\ref{fig:2axis} explores this trade-off using
example template `$\mathtt{BV}  ~\mathbf{U}~  \mathbf{G}(\mathtt{SV}
\rightarrow \mathtt{RR}$)' which was selected for evaluation to use multiple LTL 
operators and event labels.

\subsection{Example Exploit}
\label{sec:exploit}
In the Mor1kx processor, the exception, output, and basic status registers must
be equal until initialization completes. The first property in Table~\ref{tab:untilprops}
 captures this requirement. The lowest order bit of the basic status
register indicates whether the processor is the supervisor bit and a low user
changing this bit would constitute a privilege escalation. After
initialization the basic status register holds the current status unless an exception has occurred, 
in which case the status register is saved to the exception status register. I
insert a bug into the control module of the Mor1kx processor that causes this
property to be violated by the design. The inserted bug changes the initial
value of the exception status register and modifies the exception status 
register update to update all bits except
the lowest order bit, rather than the entire register. 
The system boots normally and appears identical, other than violating this
property, until an exception occurs. As soon
as an exception occurs, the correct value of the supervisor mode bit is
lost. I exploit this bug to allow an unprivileged user to enter supervisor
mode by triggering a trivial exception, so that when control returns to the
user, its status register incorrectly grants the user supervisor privileges. 

\subsection{Number of Properties}
\label{sec:traceslength}
I evaluate Undine on three open source RISC processors: Mor1kx, OR1200, and
RISC-V. Using template (5) $\mathbf{G}((\mathtt{SV} ~\&~ \mathtt{SV})
\rightarrow \mathtt{RR}$ I mine each of the three processors until a stable set
of properties is reached.
 
\begin{figure}
  \centering
  \includegraphics[width=\columnwidth]{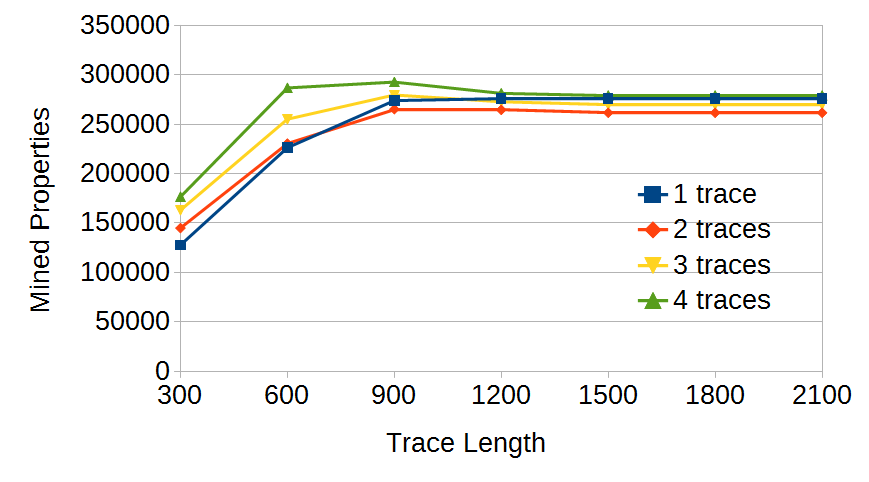}
  \caption{Steady state convergence of Undine output by trace data}
  \label{fig:steadystate}
\end{figure}

On OR1200 and mor1kx the execution traces were
\begin{enumerate}
\item arbitrary assembly code,
\item a Linux boot, 
\item the built-in test suites of the 
designs, and 
\item a bare metal hello world C program (for a system call)
\end{enumerate}
On RISC-V, I used the C program and  three benchmarks:  quicksort, 
towers, vector-vector-add. Figure~\ref{fig:steadystate} shows how the set of
properties converges to a steady state as the trace length and number of traces
increase on the RISC-V processor. The OR1200 and mor1kx processors exhibited similar trends.

The figures show the number of properties produced without
postprocessing. Table~\ref{tab:properties} shows how postprocessing reduces the final
number of properties produced
by Undine. 

\begin{table}
  \centering
\begin{tabular}{cccc}
 & Mining & Consequent & Antecedent \\
\midrule
OR1200 & 597838 & 234096 & 22 \\
\midrule
Mor1kx & 755530  & 135378 & 26 \\
\midrule
RISCV  &  278960 & 104370 & 8  \\
\midrule
\end{tabular}
  \caption{Property numbers by postprocessing stage using Template 5.}
  \label{tab:properties}
\end{table}

Runs to steady state finish in under 15 minutes.
Table~\ref{tab:timing} shows the
time it takes Undine to complete a steady state run for a given architecture, 
broken out into different stages.

\begin{table}
  \centering
\begin{tabular}{cccc}
 & Preprocessing  & Mining & Postprocessing \\
\midrule
OR1200 & 34.24 & 171.36 & 7.87 \\
\midrule
Mor1kx & 1.00  & 105.79 & 14.58 \\
\midrule
RISCV & 173.20 & 842.13 & 3.26  \\
\midrule
\end{tabular}
  \caption{Undine stage times in seconds per design.}
    \label{tab:timing}
\end{table}

\section{Limitations}

In this section, I will discuss the threats to validity for
properties produced using Undine, including false positives
and false negatives.

With regard to false positives, 
Undine utilizes template enumeration so any output
property must hold over the trace set.
Therefore, false positives are restricted to two cases: 
false positives arising from trace generation, 
and false positives arising from
misclassification, which can occur when functional properties
are classified as security properties.

With regard to false negatives, 
they fall into two cases: known and unknown.
There are three known false negatives from the evaluation
which result from failures across both the mining approach
and labelling system to address equality across bit shifts
and the introduced complexity with regards to template 
enumeration.
Reaching these properties require a either a complete redesign
(probably from template enumeration to the property inference
technique
used in other miners, such as Daikon and IODINE)
or enough processing power to make most of the Undine methodology
unnecessary.
Unknown false negatives could arise from limitations in trace
coverage, in logical specificity, within the template library,
or within the labelling system.

\subsection{Trace Reliance}

As with any specification mining technique, Undine may only determine
invariants that hold over traces. In the case of traces over buggy hardware,
discovered invariants may form a specification describing buggy behavior.
Traces may not cover all cases that can be reached by a design or even
occur during normal design operation.

For example, within the trace set, there is no Linux boot over RISC-V,
and Undine discovers only about a third as many RISC-V properties when using
Template 5. This is consistent with the RISC-V being incompletely explored by Undine,
especially
in comparison to other designs.

\subsection{Uninteresting Properties}
\label{sec:uninteresting}

Undine does
discover a number of properties prior to postprocessing 
that appear to not be relevant to any security requirements,
especially with regards to redundancy or labels not capturing
security relevance.

For example, once again consider RISC-V with regards to Template 5.
Texada finds 597838 instantiations of this labelled template
over the trace set. While these are composed into only 8 properties
during postprocessing, this simply means these properties are
unwieldy. However, 26944 of instantiations of this implication
property specify a consequent
term that is known to hold globally. There were two such terms 
of the appropriate label that were discovered from evaluating Template 1. 
These properties therefore are redundant.

Additionally, Template 5, developed
over OR1200 to find properties about 
system calls using slices of instruction registers,
overwhelmingly captures information other than whether an instruction
is a system call, a limitation of the labelling system in assessing
security relevance.

Even on OR1200, the register slices over the instruction register
produced on the order of thousands of properties while corresponding
to only tens of properties describing system calls. Ultimately, in the case
of the register slices,
the labelling system was more successful in moving mining into feasible
time frames than restricting output to only contain security properties.

Notably, this was not the case for all templates, such as
Template 9, where most of the initialization properties were interesting
in the sense that the template generated single digit numbers of properties
and manually crafting an exploit for one of these
properties was a straightforward task.

\subsection{Library Limitations}

While the template library greatly expands the number of propositional
terms that may be considered in specification mining, in testing 
it could only complete mining over
 templates of up to four terms, precluding many complex properties.
 Further, any property that does not precisely match a known template
 would not be discovered without somehow knowing to expand the library.
 Both of these are sources of false negatives.
 
\subsection{Labelling Limitations}

Relying on known properties for labelling means any register relevant
to some security property that is not known to be security relevant
would not be included in generated properties. 
Undine does generate novel properties, but does not generate properties
over novel registers.

Additionally, the labelling system required manual efforts in translating
across designs and this manual effort introduced the possibility of human error. 
Translating across designs is a significant research problem that is addressed
directly in other research efforts, particularly Transys~\citep{Zhang2020Transys}.

\subsection{Specification Logic}

Undine does not discover information flow properties or properties
requiring specification in branching time logics such
computation tree logic (CTL).

\section{Conclusion}

In this chapter I present Undine,
which automatically mines securty critical LTL properties
to create RTL specifications of processors.  Undine produces
manageable numbers of properties which, if violated, leave 
vulnerabilities over which exploits
can be readily demonstrated.  Undine runs in minutes, is usable
across different architectures and can be easily
parameterized as needed.  I also propose a labelling system
for processor events for security critical LTL properties.
Using Undine, I develop and demonstrate
 the usefulness of a library of templates in this labelling
system to secure designs against both new and known attacks.
\chapter{Isadora: Mining for Information Flow}
\label{chap:Isadora}

\section{Introduction}

Attacks targeting information flow through hardware designs
are rapidly growing in number and
severity~\citep{Chen2018SgxPectre, Evtyushkin2018BranchScope,
Kocher2018spectre, Lipp2018meltdown}.
Like Memory Sinkhole~\citep{Domas2015Sinkhole} 
and SYSRET privilege escalation attacks~\citep{Xen2012SYSRET}, 
many information flow attacks
existed in product lines for decades despite manual
verification efforts and extensive testing. What sets
information flow attacks apart is that the security properties they violate
cannot necessarily be detected by naive specification mining as the attacks 
work across multiple runs of a design. While specification mining may be extended,
it innately only generates properties for a single run of design. 
Astarte offers an automated methodology that combines Information Flow Tracking (IFT) with
specification mining to create information flow
security specifications of hardware designs.

Information Flow Tracking (IFT) is a technique to measure
flows of digital information through a hardware design by monitoring how data
propagates across elements of the design during execution. 
Recently, IFT has been demonstrated at the RTL~\citep{Ardeshiricham2017RTLIFT,Hu2018Property} 
and 
gate level~\citep{Hu2014GLIFT, Hu2016Imprecise, Becker2017GLIFT}, and has been used to
monitor implicit flows through digital side channels~\citep{Ardeshiricham2017Clepsydra}.

Existing verification engines that incorporate IFT capabilities can be used to
confirm whether a particular information flow property stated over RTL
elements holds. However, it is up to the designer to specify the full set of
desired flow behaviors, which is a difficult and time-consuming task and can
easily miss some necessary properties. To find these vulnerabilities and others an
analysis of how information flows through a design is needed. 

IFT and specification mining, together, can provide automated analysis of a
hardware design that identifies flow relations between all design
elements, including flow conditions and multi-source and multi-sink cases. The
methodology presented in this chapter is partially self correcting with respect to trace
coverage and requires no input from the designer beyond the design and a
testbench. 

To evaluate these technologies together, I developed
Isadora, a fully automatic security specification miner for information
flow properties. Isadora uses IFT
technology from Tortuga Logic's Radix-S simulation based security
verifcation engine and for specification mining uses the 
inference engine of the Daikon 
Dynamic Invariant Detector~\citep{Ernst2007Daikon}, a popular tool for mining 
specifications of programs. 
Daikon, also used for security specification mining in Chapter~\ref{chap:Astarte},
excels at specifying behavior over traces or trace slices, and was used here to
determine predicates over design state during information flows, for which it was
well suited.

Isadora demonstrates 
specification miners are capable of extracting information flow security properties 
from hardware designs. To understand security of hardware designs,
I use high level Common Weaknesses Enumerations from the database 
maintained by MITRE
as a baseline for secure design.
The results demonstrate:
\begin{itemize}
\item Isadora characterizes the flow relations between all elements of a design.
\item Isadora identifies important information flow security properties of a
  design without guidance from the designer.
\item Isadora can be used to find undesirable flows of information in the design.
\item Isadora is applicable to SoCs and CPUs.
\end{itemize} 

To measure this methodology and the usefulness of Isadora's mined
specification, I evaluated Isadora over an access control module (ACM),
a multi-controller and multi-peripheral system with a known security policy, and
a RISC-V design. I evaluated the output of Isadora versus expected
information flow policies of the design and find information flow specifications
that, if followed, protect designs from known and potential future attack 
patterns.
 
\section{Properties}

Isadora generates properties related to how information flows through a design using
trace-based analysis and specification mining. 
However, a single trace of execution cannot demonstrate how
information is flowing through the design. For example, consider a design with a
user-controlled write-ready signal ($\mathtt{WREADY}$) and an internal
write-ready signal ($\mathtt{WREADY\_int}$). While the system is undergoing a
reset cycle ($\mathtt{ARESETN} = 0$), the user should not be able to affect the
internal state of the module. It is not possible to examine any single trace of
execution and determine whether this property is being violated. 
The individual values of $\mathtt{WREADY}$, $\mathtt{WREADY\_int}$, and
$\mathtt{ARESETN}$ do not reveal how the values were calculated nor what information
was used in the calculation.

Isadora therefore leverages information flow tracking to yield traces of execution
that can be analyzed to determine how information flows through a design, and
produce output that specifies secure relations between registers.

\subsection{Example Information Flow Properties}

 While trace properties capture many important security policies, 
 some policies, such as noninterference or GMNI, for Goguen and Meseguer 
 noninterference~\citep{Goguen1982GMNI}, 
 cannot be expressed as a property of a single trace of 
 execution~\citep{Clarkson2008Hyperproperties}.  Policies over information flow, 
 including those defining security against side channel
 attacks~\citep{Lipp2018meltdown,Kocher2018spectre,Ho2018Verifying} 
 require expressing a security policy over multiple traces of execution.  
 Whereas
 trace properties are sets of traces (where a system has a property
 if its output traces fall within the set of traces making up the property), 
 information flow properties are sets of 
 trace properties, or sets of sets of traces.  Information flow properties hold
 over systems while trace properties hold for each
 discrete run of a system. 

 To explore information flow, let's consider GMNI:

 \begin{center}
 $T \in \text{GMNI} \mathrel{\mathop:}= \forall t \in T : \exists t' \in T 
  :  t \neq_H t' \land t =_L t'$
 \end{center} 

 Where $t$ and $t'$ are traces, $T$ is a set of traces that map to the set of 
 possible traces generated by some design, and relation 
 $=_L$ represents the notion of traces equal at a low level, or the
 portions of trace data visible to a low user, and 
 $=_H$ corresponding refers to the trace as view by a privileged or high user.

 Consider the case with a $\mathtt{low\_user\_data}$ signal 
 is visible to a low user and  $\mathtt{privileged\_user\_data}$ signal is 
 only visible to the privileged user. To 
 demonstrate GMNI over these signals, show that for any traces 
 $t$ there exists a trace $t'$ that differs in the value of 
 $\mathtt{privileged\_user\_data}$ 
 but which does not differ in the value of $\mathtt{low\_user\_data}$.

 \begin{center}
 $\forall t \in T : \exists t' \in T :$
 $\mathtt{low\_user\_data} = \mathtt{low\_user\_data}' \land$
 $\mathtt{privileged\_user\_data} \neq \mathtt{privileged\_user\_data}'$
 \end{center} 

 In the context of naive trace generation and specifcation mining, this is
 impossible to demonstrate because to describe the property requires 
 a comparison between multiple traces.
 There is no way from the perspective of a specification 
 miner to determine that a single trace that could be generated
 by the studied system could be added to a trace set to satisfy GMNI.
 Similarly, a single additional trace could be generated by a design but was
 not included in some studied trace set. There must be some notion
 of what possible states may exist for a design.
 
 Of note, Hyperminer~\citep{Rawat2020Hyperminer} used specification mining
 specifically to find noninterference properties of this type using a fuzzing
 technique. Doing so, Hyperminer may find many important design features.
 By way of contrast, Isadora also finds noninterference properties as discussed
 in Section~\ref{sec:propertymining}, 
 but further finds conditions
 under which interference may occur as in the example with 
 $\mathtt{WREADY}$, $\mathtt{WREADY\_int}$, and
$\mathtt{ARESETN}$ and as described in Section~\ref{sec:miningforflowconditions}. These Isadora properties correspond to the information flow
property of declassification
as described by ~\cite{Clarkson2014Temporal} wherein information may flow
between two design elements but only under defined conditions.
In the context of security specifications, 95\% of the manually defined 
security properties used for evaluating Isadora presented in 
Section~\ref{sec:refsetprops} are expressed as conditional flow properties.

\subsection{Properties with Information Flow Tracking}

 IFT offers a technqiue for measuring information flows
 between different elements of design state. 
 Isadora uses IFT at the 
 register transfer level and measures flow between registers specifically,
 rather than considering individual bits, with `registers' in this context
 referring to the Verilog notion of a register.
 Isadora may additionally be configured to consider Verilog registers and wires,
 though doing so provided no observable improvements to generated specifications
 and considerably increased trace generation costs. The Isadora methodology
 can be applied to individual bits, as the underlying information
 flow tracking used within Isadora does consider individual bits.
 However, bit level analysis would result
  in extraordinarily high trace generation costs even over smaller designs.
 
 IFT may precisely measure all digital information flows in the 
 underlying hardware, including, for example, implicit flows through 
 hardware-specific timing channels. From the perspective of specifications,
 adding IFT to a design allows the use of a 'not-flow' operator 
 (written '\texttt{=/=>}') when defining properties of the design. In the 
 case of the noninterference example:

 \begin{center}
 $\mathtt{privileged\_user\_data}$ \texttt{=/=>} $\mathtt{low\_user\_data}$
 \end{center} 

 To provide this functionality, for each signal \textit{sig} present in a design, 
 the design is instrumented in
 simulatation with a new `shadow' signal \textit{shadow\_sig}. When
 considering information flow, one or many signals may be set as the source
 from which information flow is tracked. These source signals will have
 their corresponding shadow signals set to be nonzero during trace generation,
 whereas all other shadow signals will take on an initial value of zero.
 The added instrumentation will then track flows of information from the
 source(s) through the design, toggling the relevant shadow signals from zero
 to nonzero as information flows to a given signal. In the case of GMNI over 
 $\mathtt{privileged\_user\_data}$ as a source and $\mathtt{low\_user\_data}$
 as the sink to which there should be no information flow, there is
 no interference if at a trace point the relevant IFT has tracked no information flowing
 into the $\mathtt{shadow\_low\_user\_data}$ signal:

 \begin{center}
 $\mathtt{shadow\_privileged\_user\_data} \neq 0 \land 
 \mathtt{shadow\_low\_user\_data} = 0$
 \end{center} 

 Assume in this case that no signals other than 
 $\mathtt{shadow\_privileged\_user\_data}$ are set as the initial source,
 as it could be the case that some other signal, perhaps another low user 
 data register, could flow into $\mathtt{shadow\_low\_user\_data}$.
 This is equivalent to assessing at the first trace point that for all signals 
 $s$ in the set of shadow signals in the design $S$, the starting value for 
 all signals other than $\mathtt{shadow\_privileged\_user\_data}$ is zero.
 In that case, the low value for the tracking signal
 holds over the entire trace, here given by the Linear Temporal
 Logic 'global' operator $\mathbb{G}$, that the value of 
 $\mathtt{shadow\_low\_user\_data}$ is equal to zero at every trace point.
 \begin{center}
 $(\mathbb{G}(\mathtt{shadow\_privileged\_user\_data} \neq 0)\space\land$
 \end{center}
 \begin{center} 
  $\forall s \in S \setminus \{\mathtt{shadow\_privileged\_user\_data}\} :  s = 0)\implies$
   \end{center}
 \begin{center} 
 $ \mathbb{G}(\mathtt{shadow\_low\_user\_data} = 0)$
 \end{center} 
 In this chapter, Isadora will always consider the case where a single source is specified
 along a trace. When multiple sources are considered, either multiple
 traces or multiple instances of instrumentation over a single trace can be used
 to track distinct information flows from multiple sources.
 
 \subsection{Expressing Conditions on Information Flow}
 
 Properties including 'when' keywords 
 can be expressed similarly to the logical expression of GMNI by considering
 a trace set $T$.

 \begin{center}
 $\forall t \in T : \exists t' \in T :$
$ (\mathtt{WREADY} = \mathtt{WREADY}'\implies
 \mathtt{WREADY\_wire} = \mathtt{WREADY\_wire}' ) \implies \mathtt{ARESETN} \neq 0$
 \end{center} 

 That is, if the value of $\mathtt{WREADY}$ can be determined by examining
 the value of $\mathtt{WREADY\_wire}$, then the system must not be 
 in reset. If reset is nonzero, then $\mathtt{WREADY}$ is \emph{declassified}
 and may flow to $\mathtt{WREADY\_wire}$.
 
 Figure~\ref{eq:proplinesplit} shows the equivalent expression of the write
 readiness property over a single trace using IFT.
 \begin{itemize}
 \item Line~\ref{eq:srcline} specifies the starting state of the source signal.
 \item Line~\ref{eq:nosrcline} specifies the starting state of all signals other than the source signal. 
 \item Line~\ref{eq:sinkline} specifies the sink signal.
 \item Line~\ref{eq:invline} specifies the condition on this information flow over signals
 in the original design, and is the only line not
 to refer to information flow tracking state.
 \end{itemize}

 \floatstyle{boxed}
 \restylefloat{figure}

 \begin{figure}[h]
   \centering
   \begin{subfigure}{\columnwidth}
     \centering
      \begin{equation}
 (\mathbb{G}(\mathtt{shadow\_WREADY} = 1) \land
 \label{eq:srcline}
 \end{equation}
 \begin{equation}
 \forall s \in S \setminus \{\mathtt{shadow\_WREADY}\} :  s = 0) \implies
 \label{eq:nosrcline}
 \end{equation}
 \begin{equation}
 \mathbb{G}(\mathtt{shadow\_WREADY\_wire} \neq 0 \implies
 \label{eq:sinkline}
 \end{equation}
 \begin{equation}
 \mathtt{ARESETN} \neq 0)
 \label{eq:invline}
 \end{equation}
   \end{subfigure}\\\vspace{15pt}
 \caption{A flow relation as formulated over IFT
 and original design signals.}
 \label{eq:proplinesplit}
 \end{figure}

 \floatstyle{plain}
 \restylefloat{figure}

 \subsection{Properties and Common Weakness Enumerations}
 
 This section is based on the process for creating information
 flow security properties of hardware designs put forth by
 \cite{Restuccia2021AKER} for the AKER access control module.
 An early developmental verison of AKER
  is studied by Isadora throughout this chapter.

 AKER implements the ARM Advanced eXtensible Interface (AXI) standard
 to provide secure intermodule communication for SoC designs.
 The AXI standard is a commonly used on-chip communication protocol. 
 It employs a flexible, asymmetric, synchronous interface 
 targeting high performance and low latency communications.
 An AXI architecture defines one or more controller devices accessing
 one or more peripheral devices. Different processors, accelerators,
 and other IP cores can be assigned as a controller. This allows them to
 autonomously and concurrently communicate with shared peripheral
 resources available on the SoC, e.g., a DRAM memory controller,
 ROM, and GPIOs. In this case, we will consider the AKER
 access control module specifically.

 Secure access control within SoCs are contested sites for both security
 engineers and attackers, and the
 MITRE Common Weakness Enumeration database reports a
 substantial and growing number of hardware weaknesses related to
 access control systems. While some of these are weaknesses
 expressible as trace properties and may be captured by traditional
 specification mining or other techniques, others are
 expressable only as information flows.
 In order to properly define secure information flow, Isadora must
 be able to provide specifications for hardware designs that capture
 whether a design contains a given common weakness.

 Consider this section's running example on write readiness 
 over AKER:

 \begin{center}
 $\mathtt{WREADY}$ \texttt{=/=>} 
 $\mathtt{WREADY\_wire}$ unless $(\mathtt{ARESETN}$ $\neq 0)$
 \end{center} 

 This uses the 'not-flow' operator with the addition of the 'unless' keyword to
 specify that the flow in question constitutes a violation of secure behavior
 only under certain conditions,
 in this case unless the $\mathtt{ARESETN}$ signal is not equal to zero, which means
 AKER is not undergoing reset. 
 This property was manually specified as a security property by the designers
 of AKER.
 Additionally, this flow relation partially implements secure behavior
 with respect to CWE 1272, on leaking
 sensitive information during power state transitions, for the 
 $\mathtt{WREADY}$ signal. In this case, AKER is interfacing with some
 peripheral, and the $\mathtt{WREADY}$ 
 signal is a peripheral visible register while the $\mathtt{WREADY\_wire}$
 is an internal register to AKER. The peripheral should
 not be able to interface with AKER while AKER is undergoing reset
 as access controls may not be configured.
 A flow in this case would constitute the propagation of potentially sensitive
 information
 without passing access control checks during a power state transition,
 the weakness described by CWE 1272.
 
 When assessing CWE 1272 more generally over register transfer level
 state, Tortuga Logic's Radix Coverage for Hardware Common Weakness 
 Enumeration (CWE) Guide
 provides lower level descriptions of CWEs alongside the high level 
 descriptions present in the CWE database. For example,
 CWE 1272 is described using the ``unless'' keyword:

 \floatstyle{boxed}
 \restylefloat{figure}
 \begin{figure}[h]
    \begin{verbatim}
{{Security-critical signals}}
=/=>
{{User-accessible signals}}
unless 
(privileged operating mode)
    \end{verbatim}
    \caption{CWE 1272 expressed over generic signals}
    \label{fig:eqothers}
\end{figure}
\floatstyle{plain}
\restylefloat{figure}

In consultation with the Coverage Guide, to demonstrate CWE relevance
for some property it suffices to find some source,
sink, and predicate fall within these generic
signal groups. While doing so remains nontrivial, it is a lower burden
than to define security generally.

\subsection{Security Properties for Isadora}
\label{sec:isadorasecurity}

Isadora is intended to discover properties that describe the intended 
information flow
of hardware designs.

Within the context of the Isadora framework,
the working definition of a security information flow 
property is an information
flow property over a design that corresponds to a high
level security requirement described in the 
Common Weaknesses Enumeration database
and the Radix Coverage Guide.
The CWE database is far from an exhaustive set of security 
requirements, but represents the collective efforts of
hardware engineering and security research communities.
It offers a baseline for weakness identification, 
mitigation, and prevention efforts.
This
formulation of security properties, especially with regard
to the Radix Coverage Guide, is described in greater detail by
 ~\cite{Restuccia2021AKER} specifically for access control.

Some properties produced by Isadora are classified as functional
properties under this definition, which is discussed further 
in Section~\ref{sec:functionalpropsisadora}.
 
\subsection{Properties in Implementation}
\label{sec:sampleoutput}

\floatstyle{boxed}
\restylefloat{figure}

\begin{figure}[h]
    \begin{verbatim}
case 154: 2_121_250_379_543
	_src_ in {w_base_addr_wire, M_AXI_AWREADY_wire, AW_CH_DIS, 
	w_max_outs_wire, AW_ILLEGAL_REQ, w_num_trans_wire, AW_STATE, 
	AW_CH_EN}  
	=/=> 
	_snk_ in {M_AXI_WDATA}
	unless
0 != _inv_ in {ADDR_LSB, ARESETN, M_AXI_ARBURST_wire, 
M_AXI_ARCACHE_wire, M_AXI_ARLEN_wire, M_AXI_ARREADY, 
M_AXI_ARSIZE_wire, M_AXI_AWBURST_wire, M_AXI_AWCACHE_wire, 
M_AXI_AWLEN_wire, M_AXI_AWREADY, M_AXI_AWSIZE_wire, M_AXI_BREADY, 
M_AXI_BREADY_wire, M_AXI_WREADY, M_AXI_WREADY_wire, 
M_AXI_WSTRB_wire, OPT_MEM_ADDR_BITS, S_AXI_CTRL_BREADY, 
S_AXI_CTRL_RREADY, data_val_wire, r_burst_len_wire, r_displ_wire, 
r_max_outs_wire, r_num_trans_wire, r_phase_wire, 
w_burst_len_wire, w_displ_wire, w_max_outs_wire, 
w_num_trans_wire, w_phase_wire}
    \end{verbatim}
    \caption{An example of an Isadora property, Case 154, over AKER.}
    \label{fig:isaprop}
\end{figure}
\floatstyle{plain}
\restylefloat{figure}

To consider the output properties of Isadora, Figures~\ref{fig:isaprop}
shows an example of Isadora output, Case 154 of the 303 output
properties over AKER. 
Here the predicates shown are register equality testing versus
zero. Other predicates are
captured within the workflow but not propagated to individual properties 
formatted for output.

A visible difference between an Isadora output property and
some given information flow property, such as the running example,
is that Isadora properties may specify multiple source registers,
may consider multiple sink registers though Case 154 does not do so, 
and may contain multiple predicates specifying conditions.

Case 154 includes an example of a flow condition
between internal and peripheral visible signals in addition to specifying
other aspects of design behavior. This is similar to the 
example of write readiness, but in Case 154, the flow is from
the internal signal to the peripheral, but the predicate over 
the power state is identical. 
Of note, as in the case of write readiness,
this flow occurs exclusively within the write channel, as denoted
by the ``$\mathtt{W}$'' present in ready wire and the data register.
\begin{center}
 $\mathtt{AWREADY\_wire}$ \texttt{=/=>} 
 $\mathtt{WDATA}$ unless $(\mathtt{ARESETN}$ $\neq 0)$
 \end{center} 

Under the working definition of security properties for Isadora,
where internal signals and peripheral signals should not flow to
one another unless AKER is not undergoing a reset, this
 description of behavior
composed of a single source, single sink, and single predicate
 establishes Case 154
as a security property under the working definition.
Case 154 describes signals marked as sensitive by designers,
both labelled as such within the design using comments
and present within security properties they specified, and
differs from a designer provided property only in the specific
pairing of registers.
Additionally, the paired registers match the descriptions for CWE 1272 from
the Coverage Guide.
Case 154
and other properties compared against the working definition of security properties
for Isadora are consider further when evaluating the Isadora methodology
in Section ~\ref{sec:acmcweeval}, including an example of property
not meeting the working defition.

\section{Methodology}

\begin{figure}
\includegraphics[width=\textwidth]{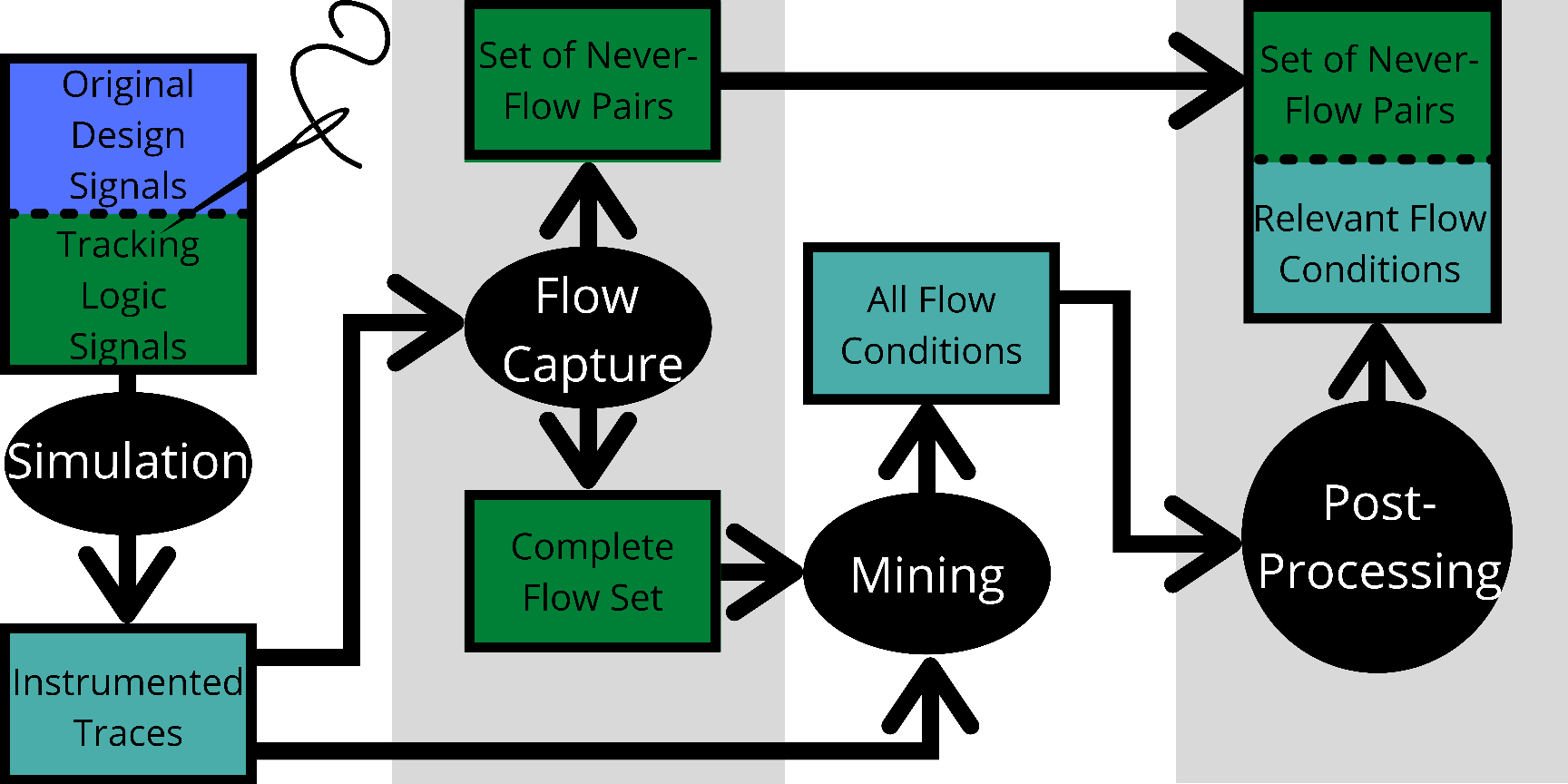}
\caption{An overview of the Isadora workflow}
\label{fig:overview}
\end{figure}

Isadora studies designs in four phases (Fig.~\ref{fig:overview}): generating traces, identifying flows, mining for
flow conditions, and postprocessing. 

The first phase
instruments the design with IFT logic and executes testbench 
over the instrumented design in simulation. 
The result is a trace of that specifies the value and tracking value
of every design signal and every time point
of execution.

In the second phase, every flow that occurred during the simulation of the
design is captured. 
This set of flows is complete: if a flow occurred between any two signals,
it will be included in this set. At the end of this phase, Isadora also
produces a set of never-flow pairs: pairs of signals between which
no information flow occurs. 

The third phase uses an inference engine to infer
predicates specifying, for every flow that occurred, the conditions under
which the flow occurred.

The final phase removes redundant and irrelevant properties from the set
of flow properties. The final set of flow properties plus the set of never-flow
properties are produced.

\subsection{Generating Traces with Information Flow Tracking}

In the first phase, the design is instrumented with IFT logic and then executed in
simulation with a testbench feeding input values to the design. 

Let $\tau_\mathtt{src} = \langle \sigma_0, \sigma_1, \ldots, \sigma_n \rangle$ be the
trace of a design instrumented to track how information flows from one signal,
$\mathtt{src}$, during execution of a testbench. The state $\sigma_i$ of
the design at time $i$ is a list of signal-value pairs describing the
current value of every original design and tracking signal in the instrumented design:
\begin{align*}
  \sigma_i = &
[(\mathtt{s}_1,v_1, v^t_1), (\mathtt{s}_2,v_2, v^t_2), \ldots,
    (\mathtt{s}_m,v_m, v^t_m)]_i.
\end{align*}

In order to distinguish the source of a tainted sink signal, each input signal
must have a separate taint label. However, tracking multiple labels is
expensive~\citep{Hu2014GLIFT}. Each added label bit doubles the state space of the
design. Isadora initially requires 10s of gigabytes of trace
data.
Therefore, Isadora takes a compositional approach. 
For each source signal, IFT instrumentation is configured to track the flow of information from only a single input signal of the
design, the $\mathtt{src}$ signal. This process is applied to each signal
in a design. 
The end result is a set of traces for design $\mathrm{D}$
and testbench $\mathrm{T}$: $\mathcal{T}_\mathrm{DT} =
\{\tau_\mathtt{src}, \tau_\mathtt{src'}, \tau_\mathtt{src''}, \ldots \}$. Each
trace in this set describes how information can flow from a single input signal to the rest
of the signals in the design. Taken together, this set of traces describes how
information flows through the design during execution of the testbench
$\mathrm{T}$.


\subsection{Identifying All Flows}
\label{sec:propertymining}

In the second phase, the set of traces are analyzed to find: 
\begin{enumerate}
\item every
pair of signals between which a flow occurs, and 
\item the
times within the trace at which each flow occurs.
\end{enumerate}
 Each trace
$\tau_\mathtt{src}$ is searched to find every state in which a
tracking signal goes from being set to 0 to being set to 1. In
other words, every tracking-signal-value pair $(\mathtt{s^t}, v^t)$ that is of the form
$(\mathtt{s^t}, 0)$ in state $\sigma_{i-1}$ and $(\mathtt{s^t},1)$ in state $\sigma_i$ is
found and the time $i$ is noted. This is stored as the tuple $(\mathtt{src}, \mathtt{s}, \{i_0, i_1, \ldots\})$, which indicates that information from
$\mathtt{src}$ reached signal $\mathtt{s}$ at all times $i \in \{i_0, i_0, \ldots\}$. 
We call this the
\emph{time-of-flow tuple}.
The tracking value of signals may be reset to zero by design events such as resets,
so the tracking value may be found to change from zero to nonzero at multiple
time points within a single trace.

Once all traces have been analyzed, the collected time-of-flow tuples 
$(\mathtt{src}, \mathtt{s}, \{i_0, i_1, \ldots\})$ are organized by time. For any
given set of times $\{i_0, i_1, \ldots\}$ there may be multiple source-sink flows that occur in the
design. Because the same testbench is used to generate every trace
$\tau_\mathtt{src}$, the timing of flows from one source $\mathtt{src}$ can be
compared to the timing of flows from a second source $\mathtt{src'}$; the value
$i$ will refer to the same point in each testbench.
At the end of this phase, the tool produces two outputs. The first is a
set of the flows through the design and the time at which they occur:
\begin{align}
S_\mathrm{flows} =  [\langle\{i_0, i_1, \ldots\}:
&\{(\mathtt{src_1}, \mathtt{s_1}), (\mathtt{src_2}, \mathtt{s_2}), 
\ldots\}\rangle;
\label{eq:flowset}\\
  \langle\{i_0', i_1', \ldots\}:
  &\{(\mathtt{src_1}', \mathtt{s_1}'), (\mathtt{src_2}', \mathtt{s_2}'), 
  \ldots\}\rangle; \ldots ]\nonumber.
\end{align}
The same $\mathtt{src}$ may flow to many sinks 
$\mathtt{s} \in \{\mathtt{s_1}, \mathtt{s_2}, \ldots\}$ 
at the same times 
$i \in  \{i_0, i_1,  \ldots\}$,
and the same sink $\mathtt{s}$ may receive information from multiple sources 
$\mathtt{s} \in \{\mathtt{src_1}, \mathtt{src_2}, \ldots\}$ 
at the same times 
$i \in  \{i_0, i_1,  \ldots\}$.

The second output from this phase is a list of source-sink pairs between which
information never flows: 
\begin{equation}
  S_\mathrm{no-flow} = \{(\mathtt{src}, \mathtt{s}), (\mathtt{src'}, \mathtt{s}'), \ldots\}. \label{eq:noflowset}
\end{equation}

The pairs in this set comprise the no-flow properties of the design, and can be
trivially rewritten using a no-flow operator, 
for example $\mathtt{src}$ \texttt{=/=>} $\mathtt{s}$. 
This provides a helpful point of comparison between Isadora's property
generation and automated tools used for property verification.
When specified on a 
single source-sink pair, these no-flows are expressible in the Sentinel language
used to write information flow assertion for Radix-S, which Isadora uses to generate
traces, and which may also be used to verifying any of of these no-flows.
However, Isadora captures all no-flow properties in the design using only
one security model per source, whereas checking all pairwise flows individual
would have an exponential higher time cost of number of sources times number
of sinks (which is one less than the number of sources). Further,
Isadora captures the flow conditions.

%

\subsection{Mining for Flow Conditions}
\label{sec:miningforflowconditions}
In the third phase, Isadora finds the conditions under which
a particular flow will occur. For example, if every time $\mathtt{src}$ flows to
$\mathtt{s}$, the register $\mathtt{r}$ has the value $x$, Isadora can infer that
$(\mathtt{src} \implies \mathtt{s}) \rightarrow \mathtt{r} = x$ may be property
of the design.

Isadora uses a miner, which itself contains an inference engine, when 
reading in traces of execution to infer design behavior using
pre-defined patterns. In order to isolate the conditions for a particular
source-sink flow, Isadora uses $S_\mathrm{flows}$ to find all the
trace times $i$ at which information flows from $\mathtt{src}$ to $\mathtt{s}$ during
execution of the testbench. The trace $\tau_\mathtt{orig}$ containing the state
of the original design is then sliced to
produce a set of two-clock-cycle trace slices, one for each time $i$. Consider
 time-of-flow tuple $(\mathtt{src}, \mathtt{s}, [i, j, k, \ldots])$, 
which as a notational convenience here uses distinct letters to denote time
points rather than subscripts for clarity in the
following expression. Given this tuple, Isadora
will produce the trace slices $\langle \sigma_{i-1}, \sigma_i \rangle, \langle
\sigma_{j-1}, \sigma_{j} \rangle, \langle \sigma_{k-1}, \sigma_{k}
\rangle$. These trace slices include only the signals of the original design,
all tracking logic and shadow signals are pruned.

The trace slices for a particular source-sink
pair are passed into the miner which infers 
the predicates capturing flow conditions
for Isadora properties.
These predicates match one of five
patterns:
\begin{align*}
  \mathtt{r} &\in \{x, y, z\},\\
  A\mathtt{r}_1 + B\mathtt{r}_2 + C &= 0,\\
  \mathtt{r}_1 &= \mathtt{r}_2,\\
  \mathtt{r}_1 &\neq \mathtt{r}_2,\\
  \mathtt{r} &= \mathrm{prev}(\mathtt{r}).\\
\end{align*}
The first line indicates that signal $\mathtt{r}$ in the original design
can take on one of three values: $x$, $y$, or $z$. The second line indicates
equality to a linear combination of terms. The third line indicates that two
signals in the original design $\mathtt{r}_1$ and $\mathtt{r}_2$ must be
equal. The fourth line indicates that two registers are never equal. 
And, the fifth line indicates that a register $\mathtt{r}$ does not
change value during the two clock-cycles of any slice. 

\subsection{Postprocessing}
\label{sec:isapost}
As a final step, Isadora eliminates redundant and uninteresting information flow 
properties. When initialling evaluating Isadora, we found that predicates of the form
$\mathtt{r}_1 \neq \mathtt{r}_2$ are usually redundant with the equality predicates of
the form $\mathtt{r} = \{x, y, z\}$ because the two registers $\mathtt{r}_1$ and
$\mathtt{r}_2$ were both constants with different constant values.
We also found that predicates involving a
linear combination of terms did not produce meaningful properties that could be
mapped back to any semantic meaning within the design. In the one case a linear
combination appeared interesting, it was actually capturing a property of a 
trivial testbench used during development. Consequentially, Isadora simply 
eliminates linear combination properties.

Finally, Isadora performs additional analysis to find invariants hold over the
entire trace set by running the miner on the unsliced trace $\tau_\mathtt{orig}$. 
One such trivial example is the invariant $\mathtt{clk} = \{0,
1\}$. Isadora eliminates any predicate that is also found to be a trace set
invariant.

The final output properties are potentially multi-source to multi-sink flows
with a number of predicates, where flows within the same property 
occur at the same time and under
the same conditions. This produces comparatively few properties, which in practice
were approximately as many as the number of unique source signals, and avoids
redundant information. Along with the no-flow properties output
directly from the second mining phase in Section~\ref{sec:propertymining}, these
conditional flows
comprise the information flow properties produced by Isadora.

This provides the second helpful point of comparison between Isadora's property
generation and automated tools used for property verification.
Using these properties from Isadora, by selecting a single source to single case
within some property, assertions may be written of the form 
$(\mathtt{src} \texttt{==>} \mathtt{s}) \implies \mathrm{Inv}$, where 
` $\texttt{==>}$ ' denotes information flow and `$\implies$' denotes logical
implication. This represents a common template used in verification.
This process is similar to that described in Section~\ref{sec:sampleoutput}
\section{Implementation}


Isadora uses the Radix-S simulation-based
security verification technology to generate IFT logic for a
hardware design, the Questa Advanced Simulator to simulate the instrumented
design and generate traces, and the Daikon~\cite{Ernst2007Daikon}
invariant miner to find flow conditions. A Python
script manages the complete workflow and implements flow analysis and 
post-processing. 


\subsection{Generating Traces}



Traces are generated for all signals within a
design. An automated utility, implemented in Python,
 identifies every signal within a design and
configures Tortuga Logic's Radix-S to build the separate IFT logic for each of these
registers. Isadora runs Tortuga in 
exploration mode, which omits cone of influence analysis, and track flows
to all design state using the \texttt{\$all\_outputs} variable. 
The resulting instrumented designs are simulated in QuestaSim over a
testbench (see Evaluation, Sec.~\ref{sec:evaluation}) to produce a trace of
execution.

\subsection{Identifying Flows}
The second phase is implemented as a Python tool that reads in the traces generated by
QuestaSim and produces the set of no-flow properties and the set of all
source-sink pairs along with their timing information. This phase combines the bit-level taint
tracking by Radix-S into signal-level tracking. 
Each $n$-bit signal in the original design is then
tracked by a 1-bit shadow signal, which will be set to 1 at the first point in
the trace that any of the component
$n$ shadow bits where set.

\subsection{Mining Flow Conditions}
\label{sec:propertyminingimpl}
The third phase is built on top of the Daikon Dynamic Invariant 
Detector~\cite{Ernst2007Daikon}, which was developed for use with software programs. 
Daikon
looks for invariants over state variables for each point in a program. For Isadora
I wrote a custom Daikon front-end in Python (411 LoC, including comments) 
that converts the trace data to be Daikon readable, 
treating the state of the design at each clock cycle as a point in a
program. The front-end also removes any unused or redundant signals and 
outputs relevant trace slices over two clock cycles as described
 in Sec.~\ref{sec:miningforflowconditions}. 

\subsection{Postprocessing}

The postprocessor is implemented as a simple Python script that interacts with
Daikon outputs through the file system. 
The postprocessor does make one additional Daikon run, over the entire trace
(rather than slices) but only considering the original design set, to 
find the set of invariants that hold globally.
Then, the the postprocessor filters Daikon output from a flow 
case based on the predicate types discussed in Section~\ref{sec:isapost} as well
as eliminating any predicates hold globally.
The postprocessor also takes the transitive closure across equalities to reduce
the number of pairwise properties and 
redundant mentions of specific registers in the output.

\section {Evaluation}
\label{sec:evaluation}

\newcommand{\refset}{designer--provided, assertion--based, 
information--flow security specification}

I evaluate the proposed methodology by assessing Isadora's ability to find information
flow security properties, especially those related to Common Weakness
Enumerations (CWEs), to answer
the following research questions:
\begin{enumerate}
    \item 
Can Isadora independently mine security properties manually developed 
by hardware designers?
    \item 
Can Isadora automatically generate properties describing CWEs over a 
design?
    \item 
Does Isadora scale well for larger designs, such as CPUs or SoCs?
\end{enumerate}

In comparison to other tools, my collaborators and I are aware of no other
tools capable of automatically generating conditions for information
flows, and these are the properties considered in Section~\ref{sec:isadoraeval2}.
~\cite{Ardeshiricham2017RTLIFT} shows register transfer level information
flow tracking may be used to verify known conditions on information flows.
Likewise, Radix-S and JasperGold are commercial tools that may
verify conditions for information flows, but require the
properties to be specified for verification.
Hyperminer ~\citep{Rawat2020Hyperminer} is a specification miner
capable of producing information flow properties, specifically
noninterference, and is discussed
in Section~\ref{sec:vshyperminer}.

\subsection{Designs}

I assessed Isadora on two designs, AKER~\citep{Restuccia2021AKER}, 
an Access Control Module (ACM),
and PicoRV32, a RISC-V CPU.
AKER wraps any AXI controller 
to implement access control by checking the validity of 
read and
write requests and rejecting those that violate a configured access
control policy. AKER was verified as secure by the designers.
PicoRV32 is a CPU core that implements the 
RISC-V RV32IMC Instruction Set, an open standard instruction set architecture 
based on established reduced instruction set computer principles.

\begin{figure}
\centering
\includegraphics[width=\columnwidth]{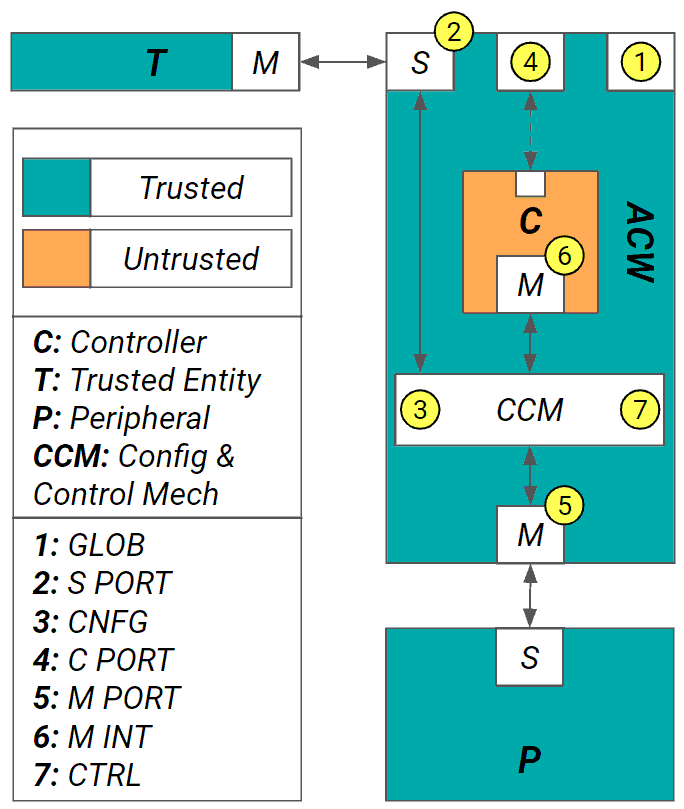}
\caption{Block diagram of the Single ACM, with the signal groups numbered}
\label{fig:blocks}
\end{figure}

\begin{figure}
\centering
\includegraphics[width=\columnwidth]{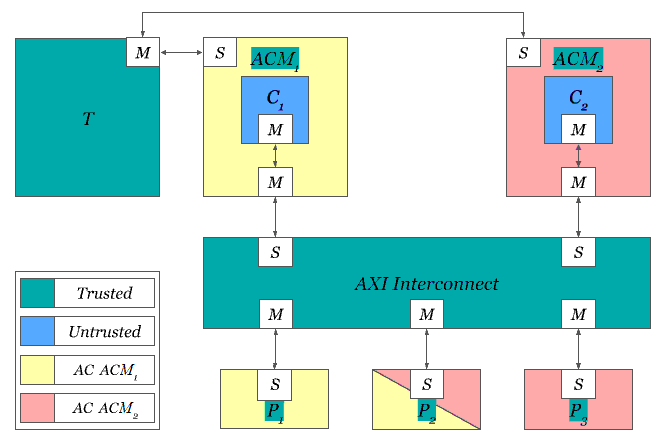}
\caption{Block diagram of the Multi ACM}
\label{fig:multiACM}
\end{figure}

I study AKER to evaluate how Isadora's properties compare to a manually developed security specification.
Isadora was tested on AKER
in two configurations: first as a stand-alone
ACM with input signals dictated by the testbench, and a second system with
two ACM-wrapped traffic generators and
three unwrapped and unprivileged modules meant to test
the AKER instances in a simulated SoC environment. I refer to these as ``Single ACM''
and ``Multi ACM'' cases, shown in Figures~\ref{fig:blocks} and~\ref{fig:multiACM}
respectively.
For both set-ups I compare Isadora's mined properties to those manually developed by hardware designers.
I use the Multi ACM case to evaluate how well Isadora scales on an SoC design.
I use the PicoRV32 to evaluate how well Isadora automatically generates properties 
describing CWEs and
to evaluate how well Isadora scales on a CPU design.

\subsection{Time Cost}

Mining was done on a system with an
Intel Core i5-6600k (3.5GHz) processor with 8 GB of RAM. Trace generation
was done on a Intel Xeon CPU E5-2640 v3 @ 2.60GHz server. Trace generation
dominated time costs, and scaled slightly worse than linear with number of 
unique signals in a design. 
Trace generation was suitable for parallelization though parallelization
was not considered in the evaluation.

The design sizes are given in Table~\ref{tab:sizetable}. 
For the Single ACM, trace generation took 
9h33m.
For the Multi ACM, trace generation exceeded
24 hours so I consider a reduced trace, which tracks sources 
for one of ACMs, though all signals are studied considered 
as sinks or in conditions.
The reduced trace was generated in 6h48m.
For PicoRV32, trace generation took 
8h35m.

\begin{table}
\centering
\begin{tabular}{ l r r r r r r r r }
\toprule
 Design & Unique& Unique & LoC & Trace & 
Trace & Daikon  & Isadora  & Miner Time\\
&  Signals & Sources &  & Cycles & GBs  & Traces& Properties & In Minutes \\
\midrule
Single ACM & 229 & 229 & 1940 & 598 & .7 & 252 & 303 & 29:51 \\
Multi ACM & 984 & 85 & 4447 & 848 & 4.3 & 378 & 160 & 8:31 \\
PicoRV32 & 181 & 181 & 3140 & 1099 & .6 & 955 & 153 & 15:09 \\
\bottomrule
  \end{tabular}
\caption{Various size measures of studied designs}
\label{tab:sizetable}	
\end{table}

\subsubsection{Theoretical Gains to Parallelization}

When parallelizing all trace generation and all case mining, Isadora could
theoretically evaluate the Single ACM case fully in less than five minutes.
Parallelizing the first phase requires a Radix-S and QuestaSim 
instance for each source register,
and each trace is generated between 120 and 180 seconds. 
Further, the trace generation time is dominated by write--to--disk, and 
performance engineering techniques could likely reduce it significantly,
such as by changing trace encoding or piping directly to later phases.
Parallelizing the second phase
requires a Python script for each source register,
and takes between 1 and 2 seconds per trace.
Parallelizing the third phase requires a Daikon
instance for each flow case, usually roughly the same number as unique sources,
and takes between 10 and 30 seconds per flow case.
The final phase, postprocessing, is also suitable for parallelization, 
but runs in under 4 seconds
on the whole design other than single Daikon instance to generate
global properties which took approximately 20 seconds including processing.  
Maximally parallelized, this gives a 
design--to--specification time of under four minutes for the single ACM and for
similarly sized designs, including PicoRV32.

\subsection{Designer Specified Security Properties}
\label{sec:refsetprops}

For the Single ACM I compared Isadora's output against 
a \refset. The specification was
 written in the Radix-S Sentinel language by applying
CWEs to the ACM using to the Radix Coverage for Hardware
Common Weakness Enumeration (CWE) Guide, 
which provides architecturally neutral
security templates for hardware CWEs.
This specification, the covered CWEs, and the results of Isadora on the 
Single ACM
are shown in Table ~\ref{tab:singleACM}. 
For each assertion
Isadora mined either a property containing the assertion 
or found both a violation and the violating conditions for each assertion.
For the provided assertions, I grouped them by source, sink, and predicate if
applicable using the signal groups in the block diagram, which were also provided
by the designers.
I reported
the observed violations to the designers who determined that the design
remained secure but a conditional flow had been incorrectly specified as 
always illegal. Isadora also found the
conditions for legality.

\begin{table}
\centering
\begin{tabular}{c|c|c|c|c|c|c}
\hline
   Source & Sink & Predicate & \# of Assert's & Result & Isadora  & CWEs  \\
   Group & Group & Group &   Provided  & & Properties &   \\
    \hline
M PORT\rule{0pt}{2.2ex} & M INT & GLOB  & 19 & \ding{51} & 2, 40,   & 1258, 1266, 1270,   \\
\cline{0-1}
\cline{4-5}
M INT\rule{0pt}{2.2ex} & M PORT &   & 19 & \ding{51} &  43, 53, & 1271, 1272, 1280\\
\cline{0-4}
\cline{7-7}
M PORT\rule{0pt}{2.2ex} & M INT & C PORT  & 19 & \ding{51} &  54, 204, & 1258, 1270, \\
\cline{0-1}
\cline{4-5}
M INT\rule{0pt}{2.2ex} & M PORT &  & 19 & \ding{51} & 214 & 1272, 1280\\
\hline
S PORT\rule{0pt}{2.2ex} & CNFG & - & 4 &  \ding{55} & 2, 6 & 1269, 1272, 1280\\
\hline
\end{tabular}
\caption{Isadora performance versus manual specification, on the Single ACM}
\label{tab:singleACM}
\end{table}

Only 9 Isadora properties, out of 303 total Isadora properties generated, 
were required to cover the designer--provided properties,
including conditions specifying violations.
The Isadora output properties may contain many source or sink signals that flow
concurrently and their corresponding
conditions. The template--based assertions considered two or three registers.
For example, on the ACM nine distinct read channel registers always flow to a 
corresponding read channel output wire at the same time, 
so Isadora outputs a single property for this design state.
This state included the reset signal and a
configuration signal both set to non-zero values, which were captured
as flow conditions, demonstrating
correct design implementation. This single
Isadora property captured 18 low level assertions related to multiple CWEs.

\subsubsection{Performance vs. Hyperminer}
\label{sec:vshyperminer}

Hyperminer has demonstrated template enumeration of flow properties with 
no predicate over SoC designs and therefore should be expected to
find any property with no specified predicate, including
 the interference between S PORTs and CNFGs in the last line of 
 Table~\ref{tab:singleACM},
which should be observable in Hyperminer 
as the absence of a noninterference property over these
registers.

While it may be the case that trace fuzzing and information flow tracking do 
not produce the same results and Hyperminer erroneously produces a noninterference
property for these registers, or that specific traces may be needed for some
properties, that is unlikely in this case.
These properties are similar to those reported by ~\cite{Rawat2020Hyperminer},
which desribes discovery of an expected decoder noninterference properties in an SoC.
Conversely, there could exist adversely affecting information flow tracking in cases
where Hyperminer does succeed, and information flow tracking, especially for RTL,
remains active research area~\citep{Ardeshiricham2017RTLIFT}.

While representing only 5\% of the properties studied in this section, these
properties contain all observed violations, so over this set of properties
Hyperminer would capture all the specification violates captured by Isadora.

By way of contrast, Isadora does additionally find negative properties 
specifying the conditions under which flows do occur, and has specificity
to discover the
remaining satisfied assertions.

These comparisons to Hyperminer similarly apply to the case of using
Isadora to study the Multi ACM.

\subsubsection{Properties of SoCs}

In the Multi ACM case, I studied
CWE 411: Unintended Proxy or Intermediary 
(`Confused Deputy'). The system contained two controllers (\textit{C}), with two 
access control modules (\textit{ACM}), a trusted entity that configured each ACM
(\textit{T}),
and three peripherals (\textit{P}). The ACMs each implemented an
access control (\textit{AC}) policy shown in Figure~\ref{fig:multiACM}
and given as:
\begin{align*}
\mathrm{AC}_1 \; \mathrm{of}\; \mathrm{ACM}_1:&\quad  R = \{P_1, P_2\},\: W = \{P_1\}\\
\mathrm{AC}_2 \; \mathrm{of}\; \mathrm{ACM}_2:&\quad  R = \{P_3\},\: W = \{P_2, P_3\}
\end{align*}
Isadora discovered legal flows from the $ACM_2$ write data to $P_3$ read and write data, 
and  $P_2$ read data. Isadora also finds an illegal flow
to $P_1$ write data. The
the $ACM_2$ to $P_1$ illegal flow has a flow condition specifying a prior flow from the 
relevant signals within $ACM_2$ to $ACM_1$. While not constituting a precise path
constraint, this captures an access control violation and suggests the 
confused deputy scenario because the flow profile from $ACM_2$ is consistent 
with this path.

These observed interference patterns are again consistent with the capabilities of
Hyperminer, which would likely find similar results.

\subsection{Automatic Property Generation}
\label{sec:isadoraeval2}

For the two designs with full trace sets, the Single ACM and PicoRV32, Isadora
generates a specification describing all information flows and their conditions
with hundreds 
of properties. To assess whether these properties are security properties, for
each design I randomly selected 10 of the 303 or 153 total properties 
(using Python random.randint) and assessed the security relevance using
CWEs.

I use CWEs as a metric to evaluate the security relevance of Isadora
output properties. 
To do so, for some design, I first determine which
CWEs apply to a design. For both the ACM and PicoRV32, I
used the Radix Coverage for Hardware Common Weakness Enumeration(CWE) 
Guide to provide a list of CWEs that
specifically apply to hardware. I considered each documented CWE for
both designs. CWEs, while design agnostic, may refer to design features not
necessarily present in the Single ACM or PicoRV32 or may not refer to
information flows. High level descriptions in multiple CWEs may
 corresponding to the same low level behavior for a design and 
 I consider these CWEs together.

Information flow hardware CWEs describe source signals, sink signals, 
and possibly condition signals.
CWEs provide high level descriptions, but Isadora targets
an RTL definition. 
To apply these high level descriptions to RTL, I group signals for a design 
by 
inspecting verilog files and grouping signals, either using designer 
notes or
manual code inspection. With the groups established, I
label every property by which group--to--group flows they contain. 
I also determine which source-sink flows could be
described in CWEs, which often correspond or even match a signal group.
I use these groups to find CWE relevant
low level signals as sources, sinks, and conditions in an Isadora property.
I also use these groups to characterize the relative frequency of 
conditional flows between different groups, which I present as heatmaps.

\subsubsection{ACM Conditional Information Flow}
\label{sec:acmcweeval}

Over the ACM I assess fourteen CWEs which I map to five plain language 
descriptions of the design features, as shown in 
Table~\ref{tab:ACMCWEgroups}.

\begin{table}
\centering
  \begin{tabular}{ l l   }
    \midrule
CWE(s) & Description  \\ \midrule
1220 & Read/write channel separation  \\ 
1221-1259-1271 & Correct initialization, reset, defaults \\
1258-1266-1270-1272 & Access controls respect operating mode  \\
1274-1283 & Anomaly registers correctly log transactions \\
1280 & Access control checks precede granting access \\ 
1267-1269-1282 & Configuration/user port separation  \\
    \midrule
  \end{tabular}
\caption{
The 14 CWEs considered for ACM}
\label{tab:ACMCWEgroups}
\end{table}

For the ACM signal groups, all registers were helpfully placed into
groups by the designer and labelled within the design. The design contained
seven distinct labelled groups:
\begin{itemize}
\item `GLOB' - Global ports
\item `S PORT' - AXI secondary (S) interface ports of the ACM
\item `C PORT' - Connections to non-AXI ports of the controller
\item `M PORT' - AXI main (M) interface ports of the ACM
\item `CNFG' - Configuration signals
\item `M INT' - AXI M interface ports of the controller
\item `CTRL' - Control logic signals
\end{itemize}

GLOB signals are clock, reset, and interrupt lines. 
S PORT represents the signals that the trusted entity \textit{T} uses to configure the ACW. 
C PORT represents the signals which are used to configure the controller \textit{C} to generate traffic for testing. 
M PORT carries traffic between the peripheral \textit{P} and the ACW’s control mechanism. 
CNFG represents the design elements which manage and store the configuration of the ACW. 
M INT carries the traffic between the ACW’s control mechanism and the controller. 
If it is legal according to the ACW’s configuration, the control mechanism will send M INT traffic to M PORT and vice versa. 
CTRL represents the design elements of the aforementioned control mechanism.

\begin{figure}
\centering
\includegraphics[width=\columnwidth]{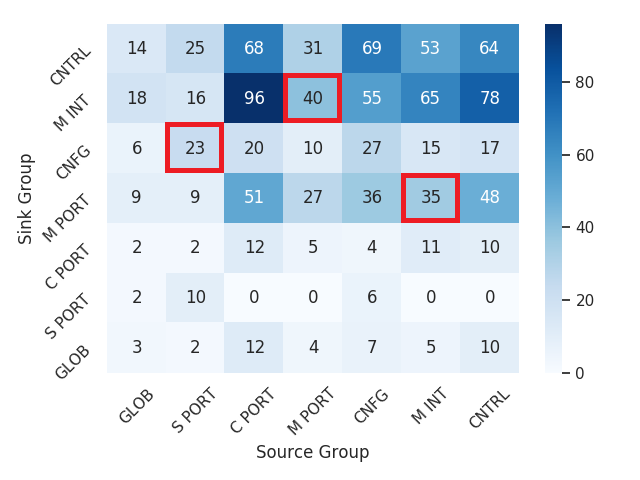}
\caption{Group-to-group conditional flow heatmap for the Single ACM.}
\label{fig:heatmap}
\end{figure}

First consider the heatmap view of the Single ACM in Figure~\ref{fig:heatmap}.
In this view, all of the designer-provided
assertions fall into just 3 of the 49 categories which are outlined
in red. Further, all of the violations 
were found with S PORT to CNFG flows, while
all satisfied assertions were flows between M INT and M PORT.
Another interesting result visible in the heatmap is the 
infrequent flows into S PORT, which is used by the trusted entity to 
program the ACM. Most of the design features should not be able to reprogram
the access control policy, so finding no flows along these cases provides
a visual representation of secure design implementation with respect to these
features.

For the ACM, all ten sampled properties encode CWE defined behavior to
prevent common weaknesses, as showin in Table~\ref{tab:CWEACM}. In this table,
the columns giving a CWE number with a `+' are referring to all the CWEs given
in a row of Table~\ref{tab:ACMCWEgroups}.
8 of 10 properties provide separation between read and write channels showing how many
flows within the design occur within these channels, which constitute the main
functionality of the ACM module. CWEs 1267-1269-1282 are not found within the 
conditional flow properties produced by Isadora as these are never flow properties,
so they are not present within the samples drawn from numbered conditional flow 
properties, but I was able to verify they are implemented
as never flows in the separate Isadora output capture these types of flows.

\begin{table*}
\begin{tabular*}{\textwidth}{rllllll}
\toprule
   \# & Description & 1220 & 1221+ & 1258+ & 1274+ & 1280 \\
\midrule
3 & Control check for first read request after reset &
 \checkmark & & \checkmark & & \checkmark \\
10 & Secure power-on &
 & \checkmark &  & &  \\
37 & Anomalies and memory control set after reset &
 \checkmark & & \checkmark &  \checkmark & \checkmark \\
96 & \textit{T} via S PORT configures ACM &
 \checkmark & & &  \checkmark & \checkmark \\
106 & Interrupts respect channel separation &
 \checkmark & & & & \\
154 & Base address not visible to \textit{P} during  reset &
  & & \checkmark & & \\
163 & Write transaction legality flows to \textit{P}  &
\checkmark  & &  & & \\
227 & Write channel anomaly register updates &
\checkmark  & &  & \checkmark & \\
239 & Write validity respects channel separation, reset &
\checkmark  & &   \checkmark & & \\
252 & Read validity respects channel separation, reset &
\checkmark  & &   \checkmark & & \\
\bottomrule
\end{tabular*}
\caption{Sampled Isadora properties on Single ACM}
\label{tab:CWEACM}
\end{table*}

\subsubsection{PicoRV32 Conditional Information Flow}

Over PicoRV32 I assess eighteen CWEs which I map to seven plain language
descriptions of the design features, as shown in 
Table~\ref{tab:RISCVCWEgroups}.

\begin{table}
\centering
  \begin{tabular}{ l l  }
    \midrule
CWE(s) & Description  \\ \midrule
276-1221-1271 & Correct initialization, reset, defaults \\ 
440-1234-1280-1299 &  Memory accesses pass validity checks \\
1190 & No flows from memory to outputs prior to reset \\
1191-1243-1244-1258-1295-1313 & Debug signals flow to no other signal groups \\ 
1245 & Correct hardware implementation of state machine \\ 
1252-1254-1264 & Data and control separation \\ 
    \midrule
  \end{tabular}
\caption{
The 18 CWEs considered for PicoRV32}
\label{tab:RISCVCWEgroups}
\end{table}

PicoRV32 had no designer specified signal groups so I used descriptive
comments regarding code sections,
register names, and code inspection to group all signals. I use
lower case names to denote these groups were not defined by the designer.
\begin{itemize}
\item `out' - Output registers
\item `int' - Internal registers
\item `mem' - Memory interface
\item `ins' - Instruction registers
\item `dec' - Decoder
\item `dbg' - Debug signals and state
\item `msm' - Main state machine
\end{itemize}
The memory interface and the main state machine were indicated
by the designer. The instruction registers, the decoder, and debug all appeared under one disproportionately large section described
as the instruction decoder. Debug was grouped by name after
manual analysis found registers in this region prefixed with `dbg\_',
`q\_', or `cached\_' to interact with and only with one another.
Instruction registers prefixed
`instr\_' all operate similarly to each other
and differently than the remaining decoder signals, which were placed
in the main decoder group. 
Internal signals were
the remaining unlabelled signals that appeared early within the design, such
as program and cycle counters and interrupt signals, and the output registers
were all signals declared as output registers.

\begin{figure}
 \centering
 \includegraphics[width=\columnwidth]{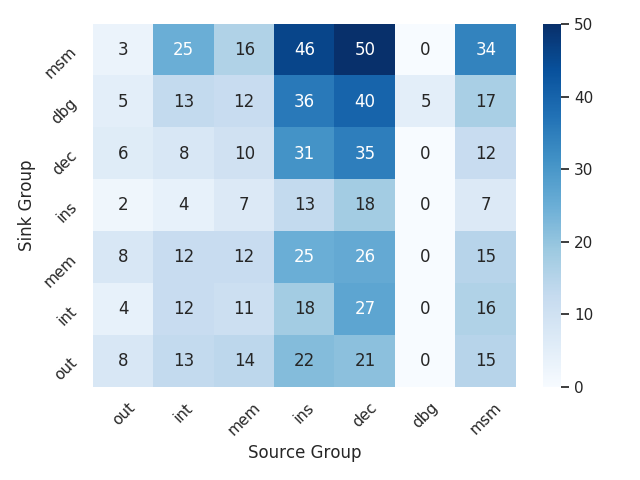}
 \caption{Group-to-group conditional flow heatmap for PicoRV32. }
 \label{fig:heatmapr5}
 \end{figure}

First consider the heatmap view of PicoRV32 in Figure~\ref{fig:heatmapr5}.
An interesting result visible in the heatmap is the flow isolation from debug
signals to the rest of the design. Many exploits, both known and anticipated
target debug  information leakage, and this entire class of weakness is shown
to be absent within the studied design at a glance.

For PicoRV32 I find eight of ten sampled properties encode CWE defined behavior to
prevent common weaknesses. I present these results in Table~\ref{tab:CWERISCV}. 
In this table,
the columns giving a CWE number with a `+' are referring to all the CWEs given
in a row of Table~\ref{tab:RISCVCWEgroups}.
The remaining two Isadora properties were single source or single sink
properties representing a simple logical combination inside the decoder, and 
captured only functional correctness. 
CWEs 276-1221-1271 are not found within the 
conditional flow properties produced by Isadora as only 3 of the 153 properties 
defined flows prior to reset and none were sampled.

\begin{table*}
\begin{tabular*}{\textwidth}{rlllllll}
\toprule
   \# & Description & 276+ & 440+ & 1190 & 1191+ & 1245 & 1252+ \\
\midrule
1 & No decoder leakage via debug &
 & &  & \checkmark & \\
16 & Instructions update  state machine &
 & \checkmark &  & & \checkmark \\
30 & Decoder updates state machine &
 & \checkmark \\
47 & No state machine leakage via debug &
 & & &  \checkmark & \\
52 & SLT updates state machine &
& & & &  \checkmark \\
66 & Handling of jump and load &
  & & \checkmark & \checkmark & & \checkmark \\
79 & Loads update state machine &
 &  &  & & \checkmark \\
113 & Decoder internal update &
  & &  & & \\
130 & Write validity respects reset &
 &  &  & & \checkmark \\
144 & Decoder internal update &
  & &  & & \\
\bottomrule
\end{tabular*}
\caption{Sampled Isadora properties on PicoRV32}
\label{tab:CWERISCV}
\end{table*}
\section{Limitations}

In this section, I will discuss the threats to validity for
properties produced using Isadora, including false positives
and false negatives.

False positives may be introduced by
 insufficient trace coverage, by
limitations of information flow tracking, or by incorrectly classifying 
functional properties
as security properties. Sampling output properties found a 
10\% false positivity rate with respect to misclassification. This rate is discussed
in greater detail in Section~\ref{sec:functionalpropsisadora}.

With regard to false negatives, 
they fall into two cases: known and unknown.
There are no known false negatives from evaluation.
Isadora was able to capture all known register transfer level
security properties.
The sampled properties 
partially addressed
all Common Weakness Enumerations manually determined to be relevant
to studied designs, but no CWE was completely covered within the 
sampled properties, so false negatives with respect to CWEs could 
still exist if some CWE relevant flow is not specified by the overall
property set.
Unknown false negatives could arise from limitations in trace
coverage or in logical specificity, which I discuss in the following
sections.

\subsection{Trace Reliance}

As with any specification mining technique, Isadora relies on traces.
The second stage of Isadora, which extracts cases where information flow
occurs, relies on generating instrumented traces with sufficient case 
coverage to drive information flow through all
channels present in the design. 
Additionally,
the third stage of 
Isadora, which mines over the original design state
to discover flow conditions, also relies on traces. 
In the case of traces over buggy hardware,
predicates detected in this stage may form a specification describing buggy behavior.
Traces may not cover all cases that can be reached by a design or even
occur during normal design operation.

Traces may not precisely describe some design features.
For example, when considering flows between internal and peripheral signals
as in the Case 154 example from Section~\ref{sec:sampleoutput}, Isadora found
a flow condition that ARLEN\_wire and AWLEN\_wire are both set to be exactly 8 for
any flow to occur.
These registers set transaction burst size for reads and writes. For
 Case 154, which only described flows in write channels, the ARLEN\_wire value
should be irrelevant,
 and this clause within the broader property constitutes
a likely false positive.

The AWLEN\_wire is a different case. In a property specifying flows during
write channel
transcations, this register would necessarily be non-zero, and for
wrapping bursts must be a power of two (wrapping is implemented by AKER),
but manual inspection of the code provides no indication the value must be
precisely 8. Some efforts were made to manipulate this and other values 
for which similar reasoning applied, but ultimately it was difficult to
tightly define possible values for which the design could operate but were
distinct from the default test bench values for this and other signals. 

My collaborators and I are currently working to address this limitation
in collaboration with Sam Meng and Kanad Basu of UT Dallas.
They study concolic test benches which may 
drive testing coverage on the basis of defined properties,
which in this case would be those generated by Isadora.

\subsection{Functional Properties}
\label{sec:functionalpropsisadora}

Under the working definition of security property for Isadora from
Section ~\ref{sec:isadorasecurity}, Isadora does contain functional
properties in output, as shown in Table~\ref{tab:CWERISCV}.
Sampling output properties found a 
10\% false positivity rate with respect to misclassification for
the sampled properties from both designs, with 0 of 10 properties
found to be false positives over the Single ACM version of AKER, and 
2 of 10 properties found to be false positives over the PicoRV32 
RISC-V CPU.

All functional properties under the working definition 
were found in RISC-V which
I attribute primarily to differences in
design and testbench. AKER was studied using a testbench specifically intended
for security research, including validation efforts related to information flow.
Further, as an access control module, by nature much of its functionality
was relevant to secure access control. For this reason, I believe
the observed error rate over RISC-V is likely closer to what would be encountered
in common practice, especially as test bench creation is
an open research problem.

With RISC-V, a minimal test bench was used that was intended
only to run the design in an environment without access to the full
RISC-V toolchain (such as the simulation environment for instrumented
trace generation), and much of the design was devoted to behavior for
which CWEs didn't not necessarily apply, such
as logical updates during instruction decoding. 
One example of an Isadora property classified as functional,
with truncated flow conditions, is presented in Figure~\ref{fig:isafunc},
and captures a logical update to an internal decoder signal.

\floatstyle{boxed}
\restylefloat{figure}

\begin{figure}[h]
    \begin{verbatim}
case 154: 128
	_src_ in {instr_lw}  
	=/=> 
	_snk_ in {is_slti_blt_slt, is_sltiu_bltu_sltu}
	unless
0 == _r_ in {alu_eq, alu_shl, alu_shr, alu_wait, alu_wait_2, ... }
0 != _r_ in {alu_add_sub, alu_lts, alu_ltu, alu_out_q, ... }

    \end{verbatim}
    \caption{An example of an Isadora property, Case 144, over RISC-V.}
    \label{fig:isafunc}
\end{figure}
\floatstyle{plain}
\restylefloat{figure}

\subsection{Measuring Interference}

Isadora assumes the correctness of the information flow tracking used
in trace generation. This is not necessarily reliant on Tortuga
Logic, and Isadora was able to use 
information flow tracking technologies
developed by Armaiti Ardeshiricham then
of UC San Diego and her collaborators in the Kastner Research Group 
to similar results in early testing, including over an AES module.

As an alternative, Isadora could perhaps be configured to interface
with Hyperminer or JasperGold to detect information flows, but 
Isadora uses timing of specific information flows.
Hyperminer finds whether an information flow occurs within a trace
set rather
 than at some trace point, and
JasperGold as a formal tool finds flows within a design rather
 than at some trace point.
 Extending either technique to study trace points would 
 likely require considerable effort and
innovation.

Nevertheless, JasperGold offers a formal method for verifying 
Isadora output properties, and could provide insight
into false positives, an immediate further
research direction.

\subsection{Specification Logic}

Isadora does not define temporal properties beyond a single delay slot
incorporated in the trace slices of length two. However,
manual examination of output properties suggests information flow
patterns during initialization,
which is the first 4 cycles for the ACM and first 80 for RISC-V, 
are highly dissimilar to latter flows.
During initialization, Isadora discovers 
flows with conditions referencing registers with unknown
states (given as `x' in Verilog and encoded as `-1' within Isadora 
to use integer encoding, a performance optimization). 
Isadora also finds 
concurrent flows between pairs of registers for which no concurrent flows occur
after reset.
By examining commingled trace slices during
and after initialization, the generated predicates may be insufficiently
precise to capture secure behavior related to this boundary. 
Isadora could likely produce more descriptive properties 
using temporal conditions on information flows, such as the
temporal property discussed in Section ~\ref{sec:exploit}.

This limitation on specifying conditions extends to all logical
expressions not produced in the third
stage of Isadora, such as implications. 
Applying the techniques developed in Chapter~\ref{chap:Astarte}
to produce more descriptive properties would offer one
approach to addressing these shortcomings.
\section{Conclusion}

I developed and implemented a methodology for creating information flow specifications of hardware designs in Isadora.
By combining information flow tracking and specification mining, Isadora is able to produce information flow properties of a design without prior knowledge of security agreements or specifications. 
Isadora characterizes the flow relations between all elements of a design and identifies important information flow security properties of SoCs and CPUs according to Common Weakness Enumerations.
 %
\chapter{Conclusion}
\label{chap:conc}

This work demonstrated specification mining can solve vital challenges for
securing hardware designs.
To demonstrate security specification mining on CISC architectures, I created
Astarte, an ISA level security property miner that partitions the x86 design 
state space along control signals that govern secure behavior of the processor.
security properties comparable to those produced by human experts.
The output specification contains properties determined to be security relevant
by manual review of design documentation, as well as additional properties
capturing correct design behavior with respect to known historical attacks, such 
as Memory Sinkhole.

To demonstrate security specification mining of temporal properties, I created
Undine, a register transer level security property miner that uses security 
specific LTL templates for specification detection. 
The output is automatically generated properties addressing
hardware vulnerabilities that can be defined before, after, and across system 
state transitions on RISC CPU designs.

To demonstrate security specification mining of information flow
properties, I created Isadora, an RTL level security property miner that uses
information flow tracking on designs to mining information flow specifications.
The output gives the information flow relation between all design elements, 
specifying whether flow may occur between two elements and, if so, under what
design conditions.
This specification provided coverage over designer--provided sets of security
properties and produced output properties that captured CWEs over multiple
designs, including an access control module, an SoC design, and a RISC-V CPU.

\clearpage
\phantomsection

{\def\chapter*#1{} 
\begin{singlespace}
\addcontentsline{toc}{chapter}{REFERENCES}
\begin{center}
\textbf{REFERENCES}
\vspace{17pt}
\end{center}

\bibliographystyle{apalike}
\bibliography{references}
\end{singlespace}
}

\end{document}